\documentclass[usenatbib]{mnras}

\usepackage[T1]{fontenc}
\usepackage{ae,aecompl}
\usepackage[dvipsnames]{xcolor}
\usepackage{mathtools}

\usepackage{graphicx}	
\usepackage{amsmath}	
\usepackage{amssymb}	
\usepackage{subfigure}
\usepackage{cancel}
\usepackage{lipsum}
\usepackage{hyperref}

\title[The role of AGN in twin galaxies]
{Testing the role of AGN on the star formation and metal enrichment of ``twin galaxies''}
\author[Angthopo et al.]
{J. Angthopo$^{1}$\thanks{E-mail: james.angthopo.16@ucl.ac.uk},
I. del Moral-Castro$^{2,3}$,
I. Ferreras$^{2,3,4}$, B. Garc{\'i}a-Lorenzo$^{2,3}$ \and and
C. Ramos Almeida$^{2,3}$
\\
$^1$ Mullard Space Science Laboratory, University College London, 
Holmbury St Mary, Dorking, Surrey RH5 6NT, UK\\
$^2$ Instituto de Astrof{\'i}sica de Canarias, Calle V{\'i}a L{\'a}ctea s/n,
E38205, La Laguna, Tenerife, Spain\\
$^3$ Departamento de Astrof{\'i}sica, Universidad de La Laguna (ULL),
E-38206 La Laguna,
Tenerife, Spain\\
$^4$ Department of Physics and Astronomy, University College London,
Gower Street, London WC1E 6BT, UK\\
}

\begin{document}
\date{Draft version \today. To be submitted to MNRAS}
\pagerange{\pageref{firstpage}--\pageref{lastpage}} \pubyear{2021} 
\maketitle
\label{firstpage}


\begin{abstract}
We explore the effect of AGN activity on the star formation history of galaxies by 
analysing the stellar population properties of 
ten pairs of nearby twin galaxies -- selected as being visually similar
except for the presence of an AGN. The selection of such twin samples represents
a method to study AGN feedback, as recently proposed by del Moral Castro et al.  
We use integral field unit (IFU) data from CALIFA, 
stacked within three fixed apertures. AGN galaxies in a twin pair 
suggest more evolved stellar populations than
their non-AGN counterpart 90\% of the time, regardless of aperture
size. A comparison with a large sample from SDSS confirms that most
twins are representative of the general population, but in each twin
the differences between twin members is significant. A set of targeted line
strengths reveal the AGN member of a twin pair is older and more metal
rich than the non-AGN galaxy, suggesting AGN galaxies in our sample
may either have an earlier formation time or follow a different star
formation and chemical enrichment history. These results are discussed within two simple, contrasting hypotheses for the role
played by AGN in galaxy evolution, which can be tested in the future at a greater detail with the use of larger data sets.
\end{abstract} 



\section{Introduction}
\label{Sec:Intro}
While substantial progress has been made over
the past decades in the subject of galaxy formation and evolution,
many questions remain unanswered, due to 
the complexity of the underlying physics, and the
limitations of both observations and simulations.
The advance in
our understanding of this field has been facilitated
through the use of high quality surveys, from 
the classic imaging and spectroscopic data of 
SDSS \citep{SDSS} to the full 3D capability of IFU-based
surveys, such as, e.g., ATLAS$^{\rm3D}$ \citep{ATLAS3D}, 
CALIFA \citep{San:2012}, MaNGA \citep{Bundy:15},
or SAMI \citep{SAMI:21}, as well as through
cosmological hydrodynamical simulations, such as
EAGLE \citep{Schaye:15} or IllustrisTNG
\citep{Mari:2018, Nel:18, Springel:18, Pillepich:18}.
High quality survey data help us discern the various trends found
in galaxy properties, while the simulations, constrained by the 
observations, enable us to gain a physical interpretation of
the key mechanisms driving galaxy formation and evolution.

There have been a number of prominent discoveries in the past few decades, 
one of the key results being the presence of a bimodal distribution 
in galaxies \citep[e.g.][]{Strateva:01}. This bimodality is the result of 
a distinct presence of two main types of galaxies: 
star-forming (SF) and quiescent (Q) galaxies.
SF galaxies are in the process of formation, and at lower redshift 
they have preferentially 
lower stellar mass, appear blue in optical bands, 
have high star formation rates (SFR) and
have significant amounts of cool gas and dust. In contrast, Q galaxies
represent an evolved version of SF galaxies \citep{Faber:07},
as they are older, appear red in the visible spectrum, have little to no 
ongoing star formation and have very little cool gas and dust.
Morphologically, SF galaxies are mainly disc/spiral and Q galaxies
are mostly classified as early-types. Bimodality has been 
observed in various planes, such as the colour-magnitude \citep{GravesBimod}, 
SFR-mass \citep{Schim:07}, UVJ bicolour \citep{2009Will}, 
colour-mass \citep{Schawinski:14} and 4000\AA\ break-velocity dispersion 
\citep[][hereafter A19]{Angthopo:19} diagrams.  

The bimodal distribution can be split into three key distinct regions in
a colour-mass plane. The two main areas are the Blue Cloud (BC), 
mostly occupied by SF galaxies, and the Red Sequence (RS), where most of the
Q galaxies lie \citep{Strateva:01,Baldry:04,Wyder:2007}.  Between these two exists
a sparse region known as the Green Valley (GV). 
The GV represents an essential stage of galaxy evolution, as 
this is where transitioning galaxies are thought to reside \citep[e.g.][]{Salim:14}. 
Galaxies on the GV are affected by key physical mechanisms
associated with halting star formation, and their spectra encode information 
about the timescales related to these processes. 
It is generally assumed that galaxies evolve from the BC to the RS
through the GV, and the sparsity of the latter suggests a fast transition
\citep[][hereafter A20]{Bremer:18,Nel:18,LDG:19,Phill:19,Angthopo:20}. However,
this image is complicated as quiescent galaxies can 
go through various episodic phases of star formation, moving
them from RS to GV or even BC, if a fresh supply of 
gas drives additional stages of recent star formation
\citep{SK:07,Rejuv2010}. Galaxy evolution is
further complicated as the dominant quenching mechanism varies
depending on the stellar mass of galaxies. At low stellar mass, 
$\lesssim 10^{10.3}\,$M$_\odot$, stellar feedback 
is thought to quench star formation in a short period 
of time \citep{Math:1971,White:1991,Hayward:2017}. In contrast, 
at high stellar mass, $\gtrsim 10^{10.3}\,$M$_\odot$, 
AGN and major mergers are thought to play an important role in quenching of
star formation
\citep{Wright:18,Bari:19,Correa:19,Dashyan:19,Terrazas:2020,Angthopo:21}. 

Even though it has become apparent that AGN are necessary 
to quench  star formation in massive galaxies 
\citep{Croton:2006,Schaye:15,SpringTNG:17},
a detailed understanding of how
AGN operate and the exact impact they have on the host galaxies 
has yet to be achieved. 
On the one hand, the triggering of strong nuclear activity is thought
to occur through galaxy-galaxy interactions or mergers, where 
the supermassive black hole (SMBH) is supplied with
fresh gas \citep{Barn:1991,HB:14}. On
the other hand, more recent observational results have shown that such
processes can occur through secular evolution, where there is no 
strong evidence of past mergers \citep{Cist:2011, Kocevski:2012, Simmons:2013}.
Due to either of these processes, it is strongly thought
that SMBH co-evolve with its host galaxy \citep{Ho:08, 2013Kor}. 
This co-evolution occurs, as SMBHs grow by accretion of the surrounding 
gas \citep{Soltan:82}, which in turn feeds an 
energetic process. This will regulate the growth of the galaxy,
as the surrounding gas will be heated or removed to such 
extent that 
the star formation is halted \citep{JR:1998}.
There have been many results strengthening this hypothesis.
One such key result
has been the high fraction of AGN found 
in the GV, as defined by the colour-magnitude \citep{Martin:07}, 
colour-mass \citep{Lac:2020} and 4000\AA\ break-velocity 
dispersion diagram (A19). Note that the general high
fraction of AGN in the GV, shows that not only are AGN the
cause of star formation quenching, following the so called 'quasar mode', 
but they also inhibit further onset of star formation by 
preventing the cooling of gas, through 'maintenance mode' feedback \citep{Ho:08}.

Other than finding out how AGN operate, one of the key questions
pertains to their universality. The prominent hypothesis dictates that
all galaxies will experience a phase where they undergo on-off cycles
of AGN activity \citep{Schawinski:15}. An alternative hypothesis
proposes, that not all galaxies will undergo an AGN phase, making AGN
galaxies somewhat unique.  The different models should result in very
different properties of galaxies. Therefore in this paper, we make use
of a sample of twin galaxies that are expected to differ only by the
presence/absence of an AGN \citep{Ig:2020} to test these scenarios.
In Sec.~\ref{Sec:Method}, we present the data and discuss the
methodology.  Sec.~\ref{sec:result1} contrasts the properties of AGN
and non-AGN galaxies in the twin samples. Sec.~\ref{sec:Disc} presents
a physical interpretation of our results in the framework of two
simple and alternative scenarios, and also discusses the potential
caveats in the interpretation of the results. Finally
Sec.~\ref{Sec:Conclusion} summarizes the main results of this paper
and its implication for future studies. To avoid confusion, all
apertures defined in the paper are quoted {\sl by radius}, unless
specifically stated.

\section{Methodology}
\label{Sec:Method}
\subsection{CALIFA data and twin galaxy selection}
\label{sec:CALIFA}
For the selection of twins, which refers to two galaxies that are almost identical 
in overall appearance and are similar in general properties, such as mass, luminosity, 
ellipticity and morphology, we make use of the 3D optical data 
(third Data release) of the Calar Alto Legacy Integral Field Area survey 
\citep[hereafter CALIFA,][]{San:2012,San:2016}. The CALIFA survey observed $\sim 667$ galaxies 
within redshift $0.005\lesssim z \lesssim 0.03$.  The parent sample
was reduced to 404 galaxies by \cite{Men:2017},  who disregarded
those that presented signs of interaction, merging activity, or had a high
inclination angle ($i>70^{\rm o}$), for a reliable characterisation of
morphology. Furthermore, they also check for presence of 
bright stars that could contaminate the systems. 
For each target that we choose to study, the survey provides three data cubes: 
V500, V1200, and COMBO.  The V500 data cubes have low
spectral resolution, covering a wavelength range of $3745 < \lambda
< 7500\,$\AA\ with $R\sim 850$ at $\lambda \sim 5000\,$\AA. The V1200
data cubes have higher resolution but cover a narrower spectral
window, $3650 < \lambda < 4840\,$\AA\ with $R\sim1650$ at
$\lambda \sim 4500\,$\AA. The COMBO data cubes are a combination of
the high and low resolution data cubes, where the high resolution
spectra, via convolution with a smoothing kernel, are degraded to
match the lowest resolution spectra of the sample. The combined data
cubes are produced to overcome the vignetting affecting the other sets.

We give below a brief description of the selection process of twin
galaxies - differing in nuclear type. More details can be found in the 
papers presenting the
original definition \citep{Ig:2019, Ig:2020}.  We start with the selection of
the AGN sample, identified with the ratios of emission line
luminosities, applying the standard BPT classification
method \citep{BPT:81}. The lines are isolated making use of the
fitting codes {\sc pPXF} \citep{2004pPXF} and {\sc
Gandalf} \citep{Sar:2006, Fal:2006}. An AGN is selected if 
the data meet the
required criteria for a Seyfert galaxy in all four different BPT
diagrams - three defined in \citet{Kewley:01} and one defined in 
\citet{CFern:2010}, that
compare the line ratios [O{\sc III}]/H$\beta$, [N{\sc II}]/H$\alpha$,
[S{\sc II}]/H$\alpha$, [O{\sc I}]/H$\alpha$, and [O{\sc III}]/[O{\sc
II}].  The AGN galaxies are included in the sample only if they are
considered to be isolated. Following the isolation criteria detailed in \cite{Barrera:2014}, 
from the whole CALIFA sample, galaxies are discarded if they meet all 
three criteria: (i) they have neighbouring galaxies within $250\,$kpc 
(ii) they have neighbours with a systemic velocity difference smaller than
$1000\,$km\,s$^{-1}$ (iii) their SDSS $r$-band magnitude difference, with the 
neighbouring galaxy, is less than 2\,mag. The
original sample was assembled to study the resolved galactic properties
\citep{Ig:2019}. This original sample was expanded upon, where the 
differences in angular momentum was explored \citep{Ig:2020}. In addition,
only galaxies with a spiral morphology, and types Sa/SBa to Sbc/SBbc
are included, thus enforcing a ``simpler'' mass assembly
history, rejecting the effect of major mergers. This selection criteria 
yielded 19 AGN galaxies.

To find the corresponding twin galaxies to these AGN (hereafter identified
as SF galaxies\footnote{However, note that the AGN galaxies in these systems also
feature ongoing star formation. This is just a convention to identify the
different twin members.}), a control
sample of star forming galaxies were selected. For the SF galaxy to
be considered a twin, firstly they have to be isolated and have to match 
the Hubble morphology of 
the AGN galaxy. In addition, the stellar mass difference between AGN and SF
galaxies has to be $\Delta \log($M$_\star/$M$_\odot) \lesssim 0.25$\,dex, the
absolute magnitude difference  $\Delta M_r \lesssim 0.70\,$mag,
the difference in SDSS-$r$ band disc ellipticity
$\Delta \epsilon \lesssim 0.2$.  The selected candidates are then 
visually inspected to ensure similarity before they are selected as 
twins.  These criteria were imposed in \cite{Ig:2020},
however we further restrict the stellar mass difference to 
$\Delta \log($M$_\star/$M$_\odot) \lesssim 0.20$\,dex
and introduce a velocity dispersion constraint, 
where the difference in velocity dispersion in the central region, within 
3 arcsec aperture, 
$\Delta \sigma \leq 30 \,$km/s.  Note the velocity dispersion
constraint is less than the typical error of measuring
velocity dispersion in CALIFA, however this hard constraint on $\sigma$
further ensures that these twin pairs, AGN and SF, galaxies are 
as similar as possible as $\sigma$ is thought
to be one of the fundamental parameters of galaxy evolution 
\citep{Graves2010DRS, Ferreras:2019}. 
The final sample comprises 
8 AGN galaxies in total and 10 twin pairs -- note that sometimes one SF can
be associated to more than one AGN galaxy as twin. 
Tab.~\ref{tab:SigtologM} shows the twin sample. From
left to right each column shows the galaxy name with their
morphological classification, their ``status'' (AGN/SF), the twin pair
used in this work, stellar
mass and velocity dispersion, star formation rate (SFR) and their 
spatial scale.
Figs.~\ref{fig:galaxy_images_1} and ~\ref{fig:galaxy_images_2}
show the SDSS images of these twins. Finally, the sample is separated 
into four groups,
based on the properties of the evolutionary diagram, discussed below. 
Note, to treat each group as a unique set, we ensure that each of the
SF galaxies only belong to one group.

\begin{table*}
  \centering
  \caption{Sample of twin galaxies, produced by identifying
  similar galaxies, where one of the twin members has an AGN.
  Cols. 1 and 2 show the galaxy ID and the morphological
  classification \citep{Walch:2014}. 
  Col.~3 labels them as AGN or star forming (SF) \citep{Ig:2020},
  note that AGN galaxies also show signatures of ongoing star formation.
  Col.~4 identifies the twin by a number. The total stellar mass 
  \citep{Walch:2014} and central
  velocity dispersion (measured within a 3\,arcsec aperture) are shown in cols. 5 
  and 6 \citep{Ig:2020}, respectively. Finally, cols. 7 and 8 shows the 
  SFR \citep[taken from][]{CataTorr:2015} and scale. }
  \label{tab:SigtologM}
    \begin{tabular}{cccccccc} 
    \hline
    Galaxy & Morph. & Type & Twin & log\,M$_\star$/M$_\odot$ & $\sigma$ (km/s) & SFR (M$_\odot$ yr$^{-1}$) & Scale (pc/\arcsec) \\
    (1) &  (2) &  (3) &  (4) &  (5) &  (6) & (7) & (8) \\
    \hline
    \multicolumn{6}{c}{Group 1}\\
    NGC2253  & SBbc & SF  & 1   & 10.50 & 109.1 & 1.13$\pm$0.19 & 257 \\
    NGC1093  & SBbc & AGN & 1   & 10.43 & 107.4 & 0.89$\pm$0.14 & 349 \\
    NGC5947  & SBbc & SF  & 1   & 10.56 & 119.3 & 1.40$\pm$0.28 & 402 \\
    NGC6004  & SBbc & SF  & 1   & 10.63 & 100.8 & 1.06$\pm$0.18 & 301 \\
    NGC2906  & Sbc  & AGN & 3   & 10.46 & 114.0 & 0.68$\pm$0.11 & 165 \\
    NGC0001  & Sbc  & SF  & 3,4 & 10.58 & 131.9 & 3.92$\pm$1.09 & 305 \\
    NGC2916  & Sbc  & AGN & 4   & 10.64 & 149.6 & 1.90$\pm$0.33 & 276 \\
    \hline
    \multicolumn{6}{c}{Group 2}\\
    NGC2639  & Sa   & AGN & 2   & 11.09 & 210.8 & 0.57$\pm$0.12 & 247 \\
    NGC0160  & Sa   & SF  & 2,6 & 10.99 & 216.1 & 0.43$\pm$0.05 & 349 \\
    NGC7311  & Sa   & AGN & 6   & 10.96 & 206.9 & 2.06$\pm$0.56 & 310 \\
    \hline
    \multicolumn{6}{c}{Group 3}\\
    NGC7466  & Sbc  & AGN & 7   & 10.68 & 128.7 & 2.85$\pm$0.49 & 509 \\
    NGC5980  & Sbc  & SF  & 7,8 & 10.69 & 121.8 & 4.31$\pm$0.75 & 320 \\
    UGC00005 & Sbc  & AGN & 8   & 10.74 & 117.9 & 4.10$\pm$0.71 & 485 \\
    \hline
    \multicolumn{6}{c}{Group 4}\\
    NGC6394  & SBbc & AGN & 5   & 10.86 & 105.8 & 1.36$\pm$0.24 & 596 \\
    UGC12810 & SBbc & SF  & 5   & 10.81 & 114.0 & 3.62$\pm$0.93 & 543 \\
    \hline
    \end{tabular}
\end{table*}

The imposition of our selection method yields state-of-the-art ``twin''
samples. Extending this to a larger set to obtain a greater number of
``twin'' galaxies requires substantially extra work and time.
The CALIFA survey observed a total of $\sim$ 665 galaxies, from which 
we only
find 10 ``Twin'' pairs. Therefore for a more statistically robust
sample, say 100 pairs, we would require a parent sample of $\sim$ 
6650 galaxies. While 
IFU surveys such as MANGA \citep{Bundy:15} have enough galaxies for this, 
not all are characterised as homogeneously as in CALIFA.

\subsection{Nebular emission correction}
\label{sec:Spec_Correction}
Before proceeding with the analysis of the stellar populations, we
must ensure we can measure the spectral features accurately. All AGN
and SF galaxies mostly have a disc-like morphology, i.e. Sa/SBa to
Sbc/SBbc, hence these galaxies have strong emission lines. We correct
for nebular emission following the methodology outlined
in \cite{FLB:13}, where we fit each of the observed spectra with
linear superpositions of simple stellar populations from
the \citet{BC03} population synthesis models, using {\sc Starlight}
\citep{SLight}. Note the use of different synthesis models,
such as eMILES or FSPS \citep{eMILES,FSPS:2010}, produce very similar
results. For each spectrum, the best fit is subtracted from the
original one. The difference produces the nebular emission component,
where the lines are fitted with a Gaussian function.  The fitted lines
are then removed from the original spectra.  However, as all the
galaxies in the twin sample have strong emission lines, we opt not to
measure the H$\beta$ index in the analysis of populations, as the
subtraction is expected to carry substantial systematic uncertainties.

\subsection{SDSS reference sample}
\label{sec:SDSS}
This paper studies the properties of the sample of twin galaxies with respect to
their evolutionary stage, following A19. We need to define the
blue cloud (BC), green valley (GV) and red sequence (RS), with a
large sample of galaxy spectra from SDSS  \citep{2006Gunn,DR14}.
Moreover, the large sample maps the general distribution of galaxies
at low redshift, that can be taken as reference to compare the
properties of the twin AGN/SF sets.
The SDSS data correspond to galaxies with a Petrosian $r$-band 
magnitude $14.5<r_{AB}<17.7\,$. The spectral coverage of the SDSS
spectrograph spans from 3800-9200\AA\ with resolution 
$R\equiv \lambda/\Delta \lambda$ of 1500 at 3800\AA\ and
2500 at 9200\AA\ \citep{Smee:13}. In order to remove a 
substantial bias from the fixed aperture, we restrict the redshift of the sample
within $0.05\lesssim z\lesssim0.1$. Furthermore, 
for our measurements to be robust, we only select galaxies
with high signal-to-noise ratio in the $r$ band, SNr$\gtrsim10$,
leaving us with $\sim 228,000$ spectra. We make use of
the {\sc galSpecExtra} catalogue \citep{Jarle:04} to retrieve the
stellar mass, BPT classification, and foreground dust. We calculate the D$_n$(4000) 
strength using a slight variation of the definition from \cite{Balogh:99}:
\begin{equation}
    D_n(4000) = \frac{\Phi^R}{\Phi^B}, \mathrm{where\,\, \Phi^{i}} \equiv \frac{1}{\lambda_2^{i} -\lambda_1^{i}} \int^{\lambda_2^{i}}_{\lambda_1^{i}} \Phi(\lambda)d\lambda,
    \label{eq:D4k}
\end{equation}
where 
$(\lambda_1^B,\lambda_2^B,\lambda_1^R,\lambda_2^R) \equiv (3850,3950,4000,4100)\,$\AA.
The spectra are corrected for foreground dust extinction, 
adopting the standard Milky Way law \citep{CCM:89}, taking the
extinction values (A$_g$) for each galaxy from the SDSS catalogues.

We define the BC, GV and RS, following a data-driven approach, where,
for a fixed stellar mass bin, we select only star-forming (SF, BPT =
1) and quiescent (Q, BPT= $-$1) galaxies. From the distribution of
SF and Q galaxies in 4000\AA\ break strength within a bin in stellar mass, 
we produce a probability distribution function (PDF) for BC (${\cal
P_{BC}}$), and RS (${\cal P_{RS}}$), respectively -- adopting  a
Gaussian distribution in D$_n$(4000). Finally, we define the GV in the same manner,
where the peak of the GV is assumed to be at the point of intersection
between the distribution of BC and RS, such that ${\cal P}_{BC} =
{\cal P}_{RS}$. We define the width of the GV distribution as half of
the width of the RS PDF.  See A19 and A20 for full details.

\subsection{Identifying Type 2 AGN in SDSS spectra}
\label{sec:T2_AGN_Selection}

In Section \ref{sec:Stats}, we contrast the CALIFA-based sample of
twin systems with the general galaxy population from SDSS.  For 
this study, the AGN galaxies in CALIFA have been selected as type~2, 
whereas the SDSS
sample is only classified, regarding AGN activity, through a simpler
classification based on the BPT diagram, that does not allow us to
distinguish between type 1 and type 2 AGN. We apply an additional
selection criterion to those SDSS galaxies classified as having an AGN, to
remove all possible type~1 AGN. The motivation for this is derived
from recent studies indicating that type~1 and 2 AGN may not be just
explained by a difference in orientation
\citep[see, e.g.,][]{CRA:11,Villar:2014, Spinogli:21, Anam:21}, in contrast to
the unification model \citep[e.g.,][]{Anto:93}. 

Previous studies from the literature typically impose a threshold in the width of the emission
lines to discriminate between type~1 and 2 AGN. Here we follow an alternative 
approach. From the  SDSS catalogues, we select the subset of SF (BPT classification = 1)
and Seyfert AGN (BPT classification = 4) galaxies.
In both sets, we measure the equivalent width of the H$\alpha$
line from the MPA/JHU catalogue, and define a simple proxy of the 
relative amplitude of the line with respect to the continuum, as follows:
\begin{equation}
    A_{H\alpha} = \frac{\phi^{max}_{H\alpha} - {\cal C}_{H\alpha}}{{\cal C}_{H\alpha}},
    \label{eq:Amp}
\end{equation}
where $\phi^{max}_{H\alpha}$ and ${\cal C}_{H\alpha}$ denote the flux of emission and  
continuum at the location of the H$\alpha$ line ($\lambda=6564.61$\AA\  in vacuum),
respectively. We divide the SF galaxies into 
different bins according to their EW(H$\alpha$) and calculate 
the mean ($\mu_{SF}$) and standard deviation ($\sigma_{SF}$)
of the distribution of $A_{H\alpha}$ values.
These estimates are used to differentiate between type~1 and 2,
so that a Seyfert AGN galaxy is 
considered type~2 if, for a given EW(H$\alpha$) its $A_{H\alpha}$
is located within 2$\sigma_{SF}$ below the mean of the distribution of
line amplitudes for the SF subset. This criterion results in 1,574 type~1
and 5,499 type~2 AGN galaxies --  consistent with previous 
studies \citep{Villar:2014}. The second
motivation for the use of 2$\sigma_{SF}$ in the classification is 
more empirical, where we argue that for a 
given EW(H$\alpha$), if the $A_{H\alpha}$ is too small, 
it must be due to the width of the line, i.e, corresponding to a type~1 AGN. 

\begin{figure}
    \centering \includegraphics[width=\linewidth]{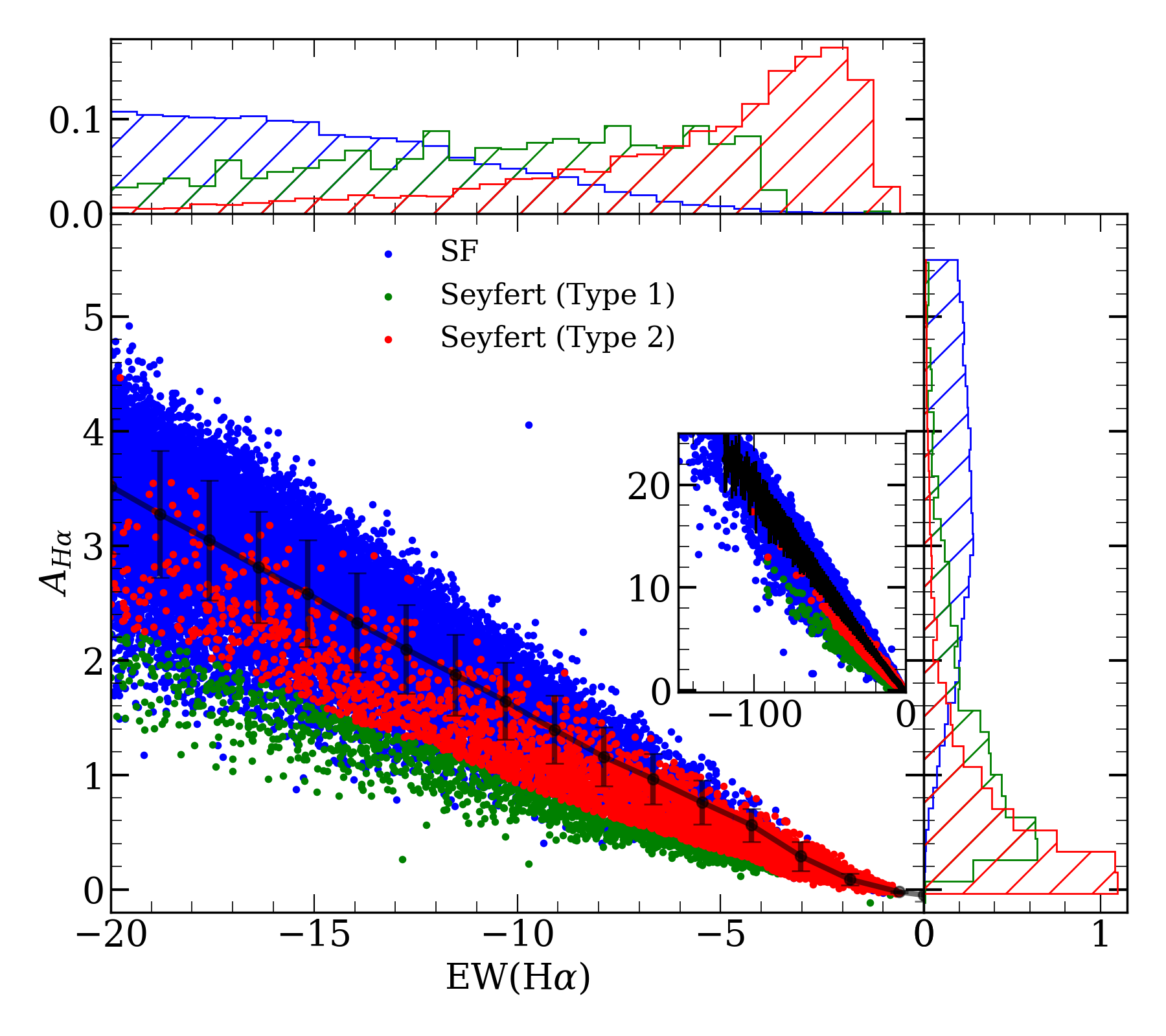}
    \caption{Distribution of star-forming and Seyfert galaxies from
    SDSS on the $A_{H\alpha}$ vs EW(H$\alpha$) plane. The data points
    correspond to galaxies selected as type~1 AGN (green), type~2 AGN
    (red), or SF (blue).  The central panel shows the region where
    most of the selected Seyfert galaxies are found, while the inset
    zooms out to show the overall distribution.  The line and points with
    (1\,$\sigma$) error bars map the distribution of SF galaxies, used
    in the classification of type~1 vs type~2 AGN (see text for
    details).  The top and right panels show the histograms of type~2,
    type~1 AGN, and SF galaxies in red, green and blue with respect to
    either EW(H$\alpha$) or $A_{H\alpha}$.}
    \label{fig:T2_sel}
\end{figure}

Fig.~\ref{fig:T2_sel} illustrates this selection criterion. 
The blue, red and green data points show the distribution of SF,
type~2 and type~1 Seyfert galaxies, respectively. 
Note, while our 
method is a crude approximation, visual inspection of the spectra of
individual galaxies confirms a low level incidence of broad line AGN galaxies in our 
type~2 definition. Additionally, Fig.~\ref{fig:Gal_Spectra} 
in Appendix~\ref{sec:spec_sel} plots the stacked spectra of type~1
and 2 AGN galaxies, binned with respect to EW(H$\alpha$), consistently showing 
that type~1 AGN have a broader H$\alpha$ line than type~2 AGN and SF
galaxies. It is possible that this procedure still suffers from some
level of cross contamination. However, due to the wide distribution of
SF galaxies, it is most likely that the method falsely flags type~2 AGN
as type~1. We argue this false classification will
only lower the number of galaxy samples therefore leaving the stellar
population properties, in general, unchanged.

\section{Stellar population differences}
\label{sec:result1}
In this section, we study various properties of the 
twin galaxy sample from integrated spectra within 
three different circular apertures. Additionally,
we test the robustness and the statistical significance of 
the results by comparing them with a larger, general sample from SDSS.
The redshift range of the SDSS and CALIFA samples is slightly different,
therefore a selection bias could be present in the comparison as:
i) the two samples correspond to different cosmic time, and/or  ii) 
we are affected by an aperture bias, as we observe different regions of 
the galaxies. We argue cosmic time is not a major issue as it does 
not vary significantly between the different redshift ranges probed by the
samples, $\lesssim 1\,$Gyr. Regarding the latter, we assess the effect 
by producing CALIFA spectra within apertures that are equivalent 
to the area covered (in physical size) by the 3\arcsec diameter of the
fibers of the original SDSS spectrograph.

\subsection{Differences in the evolutionary stage of twin galaxies}
\label{sec:TG_Evo}

\begin{figure*}
    \centering
    \includegraphics[width=\linewidth]{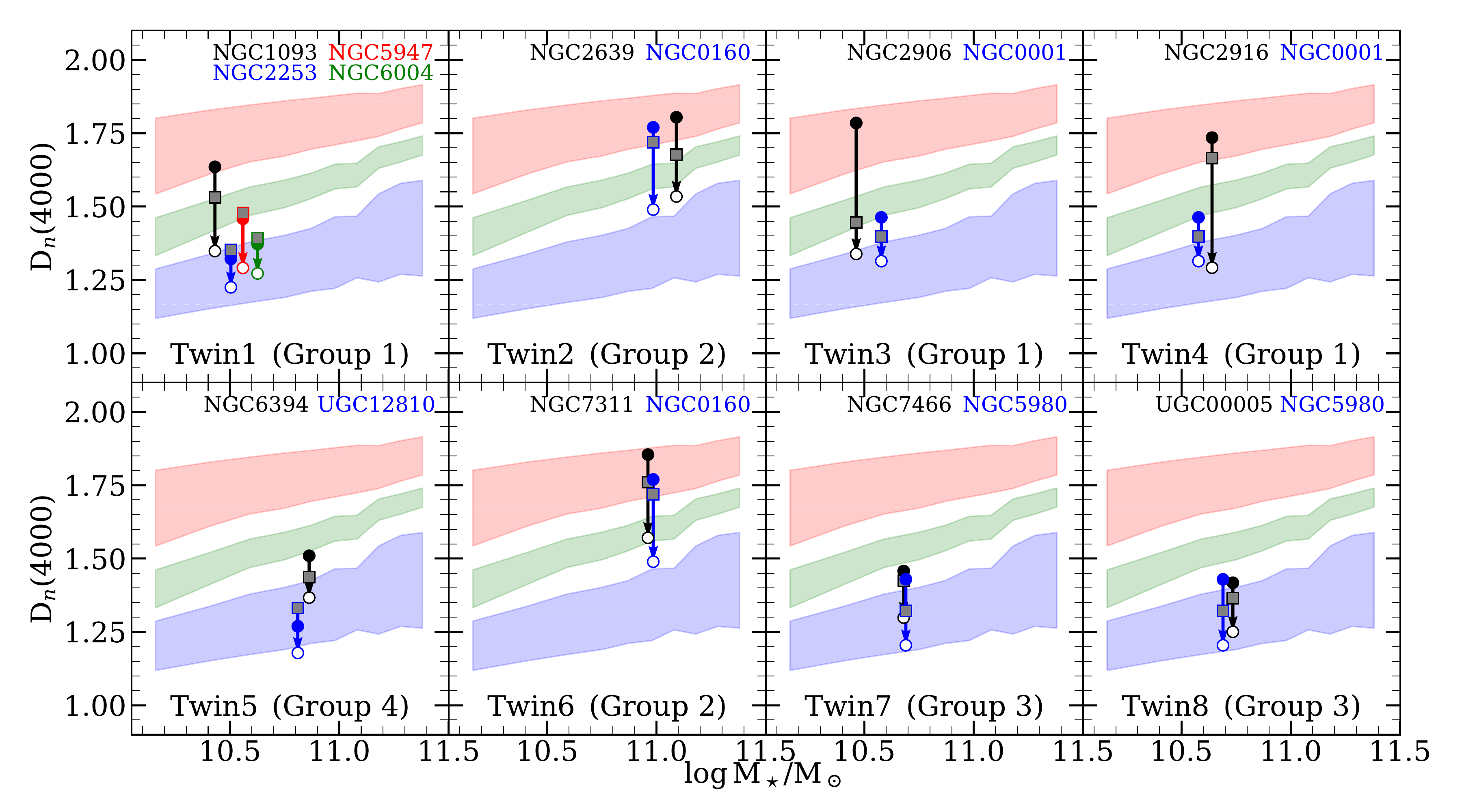}
    \caption{Distribution of twin galaxies on the 
    D$_n$(4000) vs stellar mass plane. The blue, green and
    red filled regions respectively map the BC, GV and RS  defined
    in A19 and A20. Each panel shows the twin 
    galaxy pairs, where the black circles are 
    the AGN-hosting galaxies, and the coloured circles represent
    the non-AGN (i.e. SF) counterparts, all measured within a
    1.5\arcsec aperture, similar to the PSF of CALIFA. The grey squares 
    are the results within a 2.2\,kpc 
    aperture radii, and the white circles correspond to the  largest 
    aperture in this study, R$\leq1.5$R$_{\rm eff}$. Note, the error 
    on D$_n$(4000) is calculated via Monte Carlo realisation, where due to
    high SNr on the spectra, uncertainties on measurements of 
    the index is minuscule, $\Delta$D$_n$(4000)$\leq 0.01$. 
    From the original twin
    sample of 11 systems (10 Type 2 Seyfert AGN and 1 Type 1 Seyfert AGN), 
    we reject three (see text for details) pairs.}
    \label{fig:Bimod_Min}
\end{figure*}

In order to compare the evolutionary stage of twin galaxies, we adopt
the D$_n$(4000)-stellar mass plane in Figure~\ref{fig:Bimod_Min} --
instead of the standard colour-stellar mass diagram
\citep{Schawinski:14,Bremer:18}. Each panel corresponds to a different twin.
The blue, green and red filled regions depict the BC, GV and RS 
respectively; defined by the SDSS data (A19).
The black filled circles, grey squares and unfilled circles  
show the result for measurements within three different apertures,
namely 1.5\arcsec, 2.2\,kpc, and 1.5\,R$_{\rm eff}$ (all defined by radius),
respectively - Fig.~\ref{fig:apertures} illustrates, for a single 
galaxy, the aforementioned apertures.
The first one (R$\leq$1.5\arcsec) concerns the data from the innermost region of
the galaxies, thus studying the immediate vicinity where the AGN has 
the highest impact. It also takes into account the smearing of the 
point spread function.
The R$\leq$2.2\,kpc is justified for a comparison between SDSS data and this
sample, by matching the aperture size to the physical extent of the SDSS fibre 
at the median SDSS redshift (z$\sim$0.077). Note the median redshift for 
the CALIFA survey 
is z$\sim$0.015. The last aperture (1.5\,R$_{\rm eff}$) is meant to explore most
of the galaxy, within the reach of the CALIFA survey.
Note the D$_n$(4000) measurements are carried
out on spectra that have been smoothed to a common
velocity dispersion, $\sim 220\,$km/s (i.e. approximately 
the maximum value in the sample). 
In each twin, the AGN galaxy is shown in black, while the SF counterpart(s) 
are shown in blue, green or red. The strongest trend in this sample shows that 
9 out of 10 twin pairs have an AGN with a greater 4000\AA\ break than their SF
counterparts, even amongst twin samples, 
in both their central regions (1.5\arcsec), and within 2.2\,kpc, 
indicating the importance of AGN in quenching star formation \citep{Schawinski:07, Bell:2008}. 
The case for a 2.2\,kpc aperture -- defined  to match results with 
SDSS spectra -- shows that most AGN galaxies, $\sim 71\%$, 
reside in the GV. Here we have identified galaxies as residing in GV, even if
they are slightly above BC or below RS; this is based on how we define 
GV galaxies in A19 and A20. Previous work from the literature based on 
SDSS, MaNGA and CALIFA data have also noted that a high fraction of
AGN, specifically LINER galaxies occupy the GV 
\citep{Martin:07, San:2018, Angthopo:19, Lac:2020}. In addition, both AGN and 
SF galaxies  
show a decrease in 4000\AA\ break strength radially
outward, thus indicating older stellar populations  
in the centre, suggestive of inside-out 
quenching \citep{Kelvin:18}. The only exception to 
this trend is seen in the 3 SF galaxies belonging to twin 1 and twin 5,  
where the stellar
populations within 2.2\,kpc have a greater D$_n$(4000) index
than those within 1.5\arcsec. All 4 SF galaxies are barred spiral galaxies, indicating
younger stellar populations at the most central region. Note, \cite{Ig:2019} also 
found similar results in the pilot study regarding twin galaxies. 

Using the aperture of the most central region, i.e.  1.5\arcsec, 
the twin(s) are binned into different groups (as labelled in Fig.~\ref{fig:Bimod_Min}),
based on their location in the diagram. Group 1 (G1) consists of twins 1, 3 and 4,
as their AGN galaxy resides in the RS, whereas their SF galaxies
reside in either the BC or GV. Group 2 (G2) 
consists of twins 2 and 6, since both their
AGN and SF galaxies are in the RS. 
Group 3 (G3) is made up of twins 7 and 8, as both AGN
and SF galaxies reside in the GV/BC region.
Finally, twin 5 alone is grouped in G4 as this is the only 
twin with the AGN galaxy in the GV, whereas its SF counterpart
is clearly in the BC. The grouping is robust 
as it does not depend significantly on the adopted aperture.
Using the R$\leq$2.2$\,$kpc spectra to define the groups, we find $\sim$22\% of the 
grouping would change, as twin 4 
would be on its own group and twin 5 would join twins 7 and 8. 
Similarly, using the 1.5\,R$_{\rm eff}$ aperture, the  grouping structure would 
remain the same, but twins 5, 7 and 8 would be put together.

From these groups, we aim to test whether variations in stellar
properties are caused either by the presence of an AGN, or by
the galaxies being at different stages of evolution. It is
evident for G1 that AGN galaxies are more evolved than SF galaxies
within 1.5\arcsec. Note that while the D$_n$(4000) vs stellar mass diagram
is a proxy of evolution, as galaxies are expected to transition
from the BC to RS via the GV, the actual path could be complicated by
rejuventaion events \citep{Rejuv2010}. We argue that rejuvenation should
not play an important role in our sample as (i) these galaxies are
selected to be isolated, therefore it is unlikely they have undergone 
major merging event. However recent minor mergers cannot be ruled
out and
(ii) state-of-the-art simulations, IllustrisTNG, have noted
that rejuvenation events are more prominent at high stellar mass
$\gtrsim 10^{11} M_\odot$ \citep{Nel:18}, thus above the mass range probed
by this sample.

\subsection{Contrasting with the general galaxy population (SDSS)}
\label{sec:Stats}
Fig.~\ref{fig:Bimod_Min} therefore suggests that AGN galaxies tend to
be more evolved compared to their twin SF counterparts. Previous work
noted this trend as well, where a high fraction of AGN
galaxies is found in the GV \citep{San:2018,Lac:2020}, whereas most of the SF galaxies
are located in the BC. However, our result is more focused, as it targets
sets of carefully defined pairs with very similar overall
properties except
for the presence of an AGN. Owing to the small sample size, we address now 
the statistical significance of our results, by comparing
the sample with a large, general distribution of galaxies from the SDSS.

\subsubsection{Line strengths}
\label{sec:LS_sub}
\begin{figure*}
    \centering 
    \includegraphics[width=\linewidth]{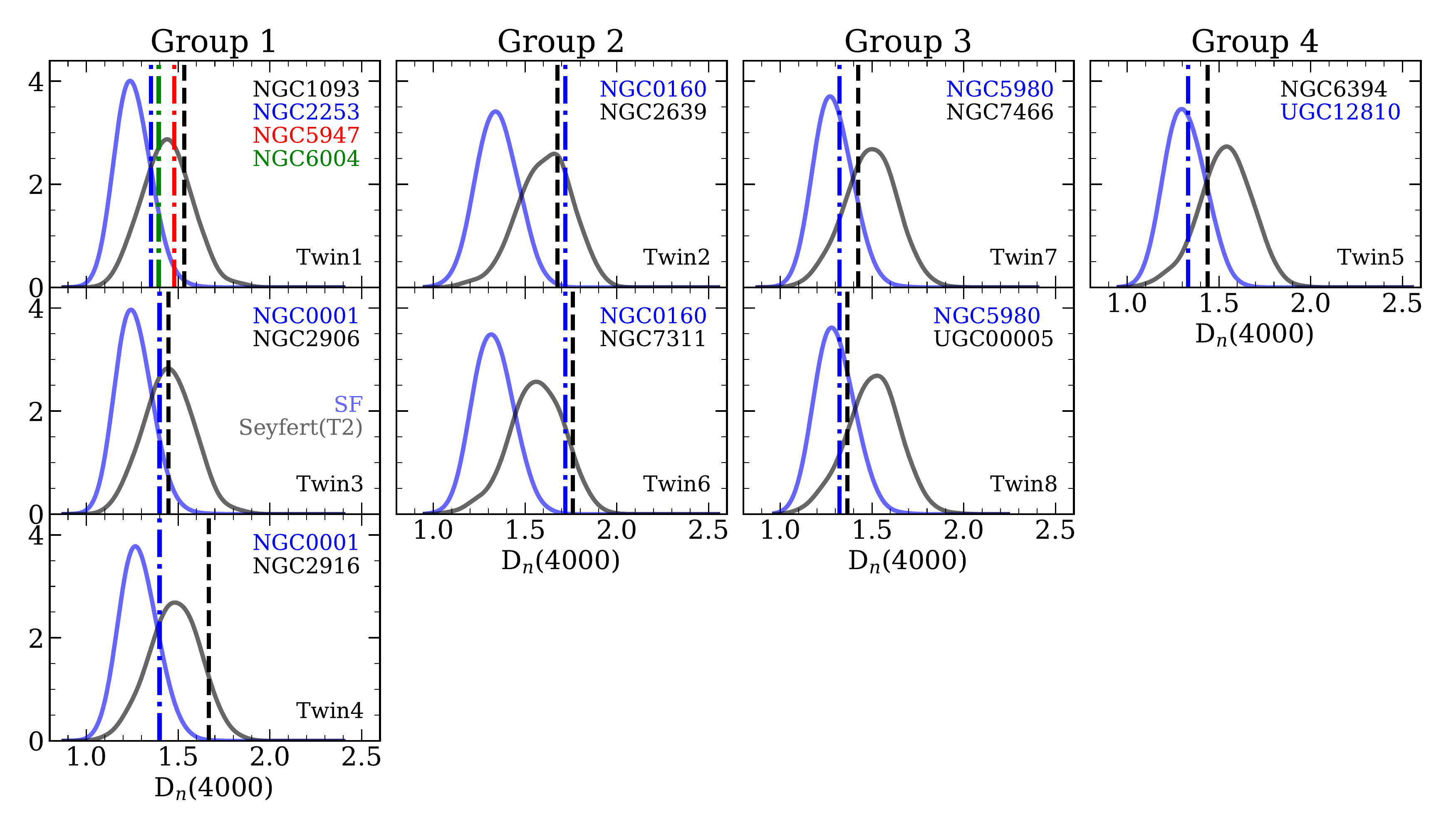}
    \caption{General distribution of D$_n$(4000).  The blue
    and grey Kernel Density Estimation (KDE) 
    histograms show the distribution of star-forming
    and type 2 Seyfert AGN from a sample of SDSS within a range of stellar mass
    compatible for each twin. The black dashed vertical line shows the D$_n$(4000) index
    for the AGN in the twins, while the dash-dotted lines, with varying colours, show the
    D$_n$(4000) index for SF twins. The twin estimates are given for a 2.2$\,$kpc aperture to
    match the SDSS sample. See text for details.}
    \label{fig:D4k_SDSS}
\end{figure*}

The differences found in Fig.~\ref{fig:Bimod_Min} are still open to
a potential sample selection effect. We need to assess 
whether the population properties of the twin galaxies
is comparable with the general sample, i.e. are these galaxies a fair
representation of their counterparts in SDSS, or are they 
statistical outliers? We compare key line strengths, D$_n$(4000),
H$\delta_A$ and [MgFe]$^\prime$ within the 2.2\,kpc aperture with respect to
the distribution of galaxies in SDSS. We select spiral galaxies from
SDSS, making use of the Galaxy Zoo catalogue \citep{Lin:08}, choosing
the spiral flag set to 1.  Additionally, we do not smooth the spectra
of the twin galaxies to $\sigma \sim 220\,$km/s, unlike in previous and
future sections, as the SDSS spectra are not smoothed either. Note,
the smoothing has a small effect on the line strength, where we
find the largest differences for [MgFe]$^\prime$, with a maximum offset of
$\sim$0.26\,dex.  Fig.~\ref{fig:D4k_SDSS} shows the distribution of
4000\AA\ break strength for each twin -- where the SDSS sample is
restricted to the same stellar mass, within
$\Delta\log$M$_\star/$M$_\odot=\pm$0.2\,dex.  The vertical dashed lines
locate the D$_n$(4000) index for the twin galaxies: black for the AGN,
coloured lines for SF systems.  The solid blue and grey 
histograms show the distribution of SDSS galaxies classified as
SF and type~2 Seyfert, respectively.

In G1, both twins 1 and 3 show the AGN galaxies close 
to the peak of the type~2 AGN distribution. In contrast, the AGN galaxy of twin~4 
is offset with respect to the peak of the Seyfert distribution, by 
$\gtrsim 1\sigma$. 
SF galaxies in all 3 twins belonging to group 1 are more representative
of the general sample, and are located 
towards the peak of the SDSS SF galaxy distribution. Note, while
the corresponding histogram for the complete sample is not shown it 
also peaks at the same location as the "SF" sample,
indicating a high absolute number of SF galaxies within the chosen
stellar mass. SF galaxies 
in G2 are located near the tail end of the SDSS SF galaxy 
distribution, whereas the AGN are neighbouring the peak of SDSS type~2
Seyfert distribution. Thus, while possible, such twin pairings are 
unlikely if extracted randomly from a larger, general sample. 
G3 and G4 show similar trends, to that of G1,
where the  4000\AA\ break of the AGN is located close to peak of 
Seyferts. For each
twin, the SF galaxies tend to have their 4000\AA\ break close
to the peak of the SDSS SF distribution. Therefore, if we were to
extend the sample to a larger survey, we would find that twins 1, 3, 4, 5, 7 and 8
are representative samples of the general population. 

\begin{figure*}
    \centering
    \includegraphics[width=\linewidth]{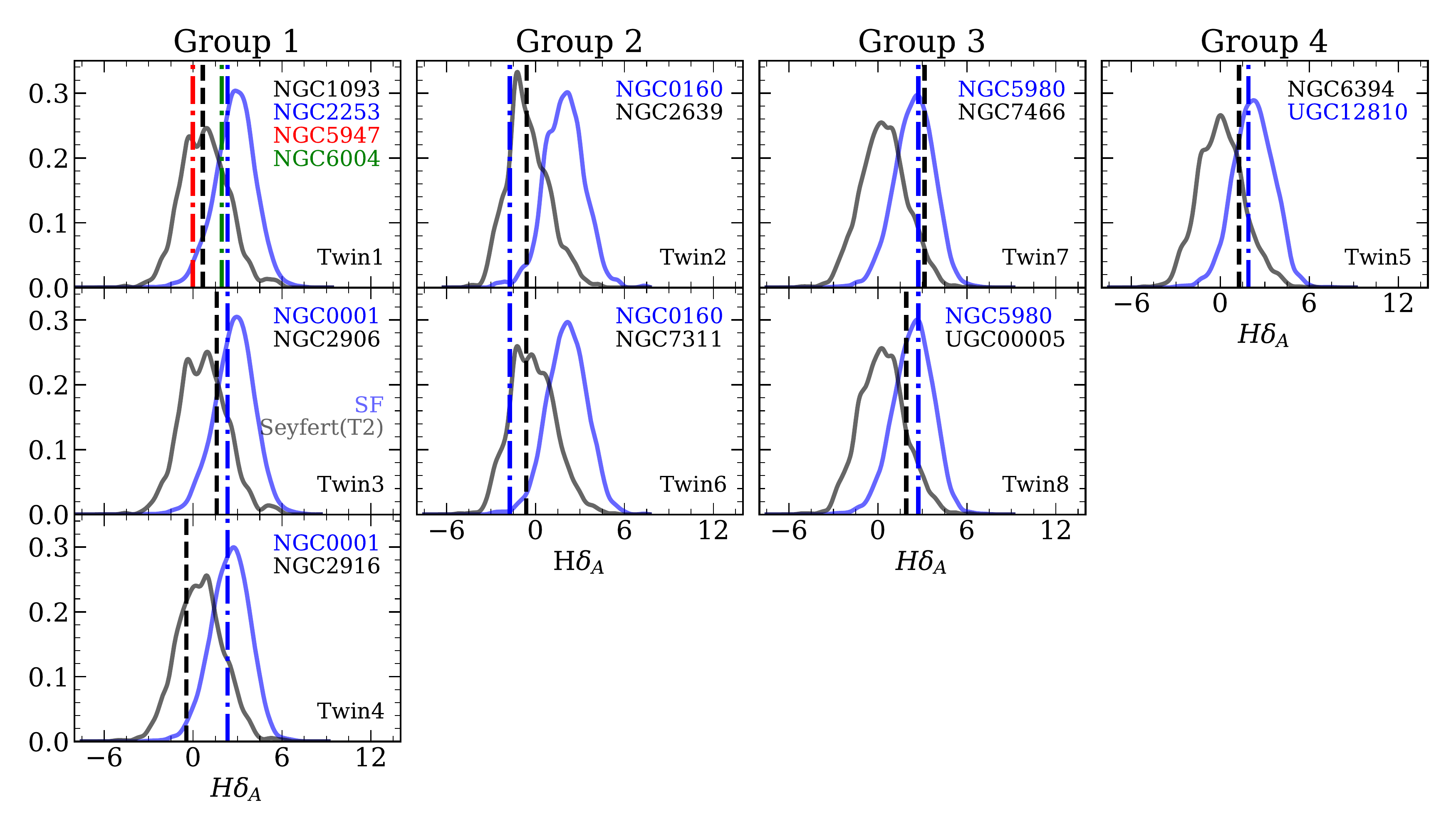}
    \caption{Same as Fig.~\ref{fig:D4k_SDSS} for the distribution of 
    H$\delta_A$ line strengths.}
    \label{fig:HdA_SDSS}
\end{figure*}

Fig.~\ref{fig:HdA_SDSS} and Fig.~\ref{fig:MgFe_SDSS} show 
the H$\delta_A$ and [MgFe]$^\prime$ distribution for AGN  and SF galaxies, with
the colour-coded vertical lines once more representing the individual measurements
of galaxies in the twin samples. 
The SDSS distributions follow the same labelling system 
as Fig.~\ref{fig:D4k_SDSS}. All twins in G1 have H$\delta_A$ 
for AGN and SF galaxies compatible with respect to the SDSS distribution, with the exception 
of NGC5947 in twin 1. Similarly to D$_n$(4000), the H$\delta_A$ strength 
of SF galaxies in G2 are at the 
tail end of the distribution, while AGN galaxies are located
closer to the peak of type~2 Seyfert AGN. 
G3 and G4 suggest a high likelihood of finding both AGN and SF galaxies
with the observed H$\delta_A$, with the exception of twin~7 
where the AGN is located $\gtrsim 1\sigma$ away from the peak of the distribution.
For G1, G3 and G4, the [MgFe]$^\prime$ index in both AGN and SF galaxies resembles 
that of the larger SDSS sample, once more with the exception of 
NGC5947. In comparison, G2 shows the twin SF galaxy to 
deviate away from the peak location by $\gtrsim 1\sigma$, whereas 
the AGN galaxies appear closer to the peak of the distribution in type~2
Seyfert systems. The inclusion of H$\delta_A$
and [MgFe]$^\prime$ shows, given their position in the 
evolutionary sequence, how alike the twin galaxies are to a larger 
parent sample.

\subsubsection{Testing the significance of the relative differences}
\label{sec:rel_diff}
We focus here on one key question: how significant is the difference
found in the D$_n$(4000) vs stellar mass plane between the twin pairs,
with respect to a random pairing of galaxies with similar mass? To
answer this,
we look at the relative difference in 4000\AA\ break strength between 
the twin samples, measured within a  2.2\,kpc aperture, and compare it with 
random pairings of galaxies with similar stellar mass (within $\Delta\log$M$_\star/$M$_\odot=\pm$0.2\,dex)
from SDSS. We define the relative difference as follows:
\begin{equation}
    \delta (\Psi) = \frac{\Psi_1 - \Psi_2}{\Psi_1},
    \label{eq:RD_vals}
\end{equation}
where $\Psi_1$ and $\Psi_2$ are the line strengths of the galaxies in each pair.

Fig.~\ref{fig:RD_D4k} shows as vertical lines the observed relative
difference in the twin sample, where the blue, red, green and orange 
lines show differences for individual twins, as labelled, while the
black vertical line shows the average of the four twins.  Note twin 1
has multiple pairs comprising 1 AGN and 3 SF galaxies, therefore we show the 
average of the three pairs.  The SDSS data are represented as a Gaussian
distribution (i.e. taking the mean and standard deviation of the SDSS
sample to produce a Gaussian equivalent). The blue distribution shows
the result when selecting random pairs from the subset of only SF galaxies,
in SDSS and the grey distribution corresponds to random pairings that include
SF and type 2 Seyfert AGN galaxies. Note these pairings can be SF-SF, 
SF-AGN or AGN-AGN. The figure is split according to stellar
mass, where the left panel shows differences for SDSS galaxies within
$10.28\lesssim \log\, \mathrm{M_\star/M_\odot}\lesssim 10.78$ and the
right panel shows differences within
$10.66\lesssim \log\, \mathrm{M_\star/M_\odot}\lesssim 11.16$. We find
all twins (except twin 4) to have relative differences within
1$\sigma$ (2$\sigma$) of the SDSS relative difference distribution for
both SF and AGN galaxies. Note, while the variation seems to be
statistically small, the relative difference is generally positive, 
suggesting the results 
found in Fig.~\ref{fig:RD_D4k} indicate physical differences between 
the evolutionary stage of twin galaxies, so that galaxies hosting an AGN
have stronger 4000\AA\ break strengths than their SF counterparts - 
even when considering "twin" galaxies. This paper relies on a rather small set
of twin galaxies as  a proof of concept. Future studies, with larger
data sets, will allow us to produce more conclusive results. However, 
the present sample already shows the high diversity in twin pais, which
rules out simple AGN feedback models.

\subsection{Inner vs Outer regions}
\label{sec:InOut_Diff}
We consider now the difference in line 
strength between the smallest aperture, 1.5\arcsec, and the largest one,
extending to 1.5\,R$_{\rm eff}$, by comparing the 
age-sensitive indices,  D$_n$(4000) and H$\delta_A$, as well as the 
metallicity-sensitive index, [MgFe]$^\prime$. We define the 
absolute difference between line strengths with respect to aperture as: 
\begin{equation}
    \Delta \Psi = \Psi(R\leq 1.5\arcsec) - \Psi(R\leq 1.5R_{\rm eff}),  
    \label{eq:rad_diff}
\end{equation}
where $\Psi$ represents the different indices --  either
D$_n$(4000), H$\delta_A$ or [MgFe]$^\prime$.

\begin{figure*}
    \centering
    \includegraphics[width=\linewidth]{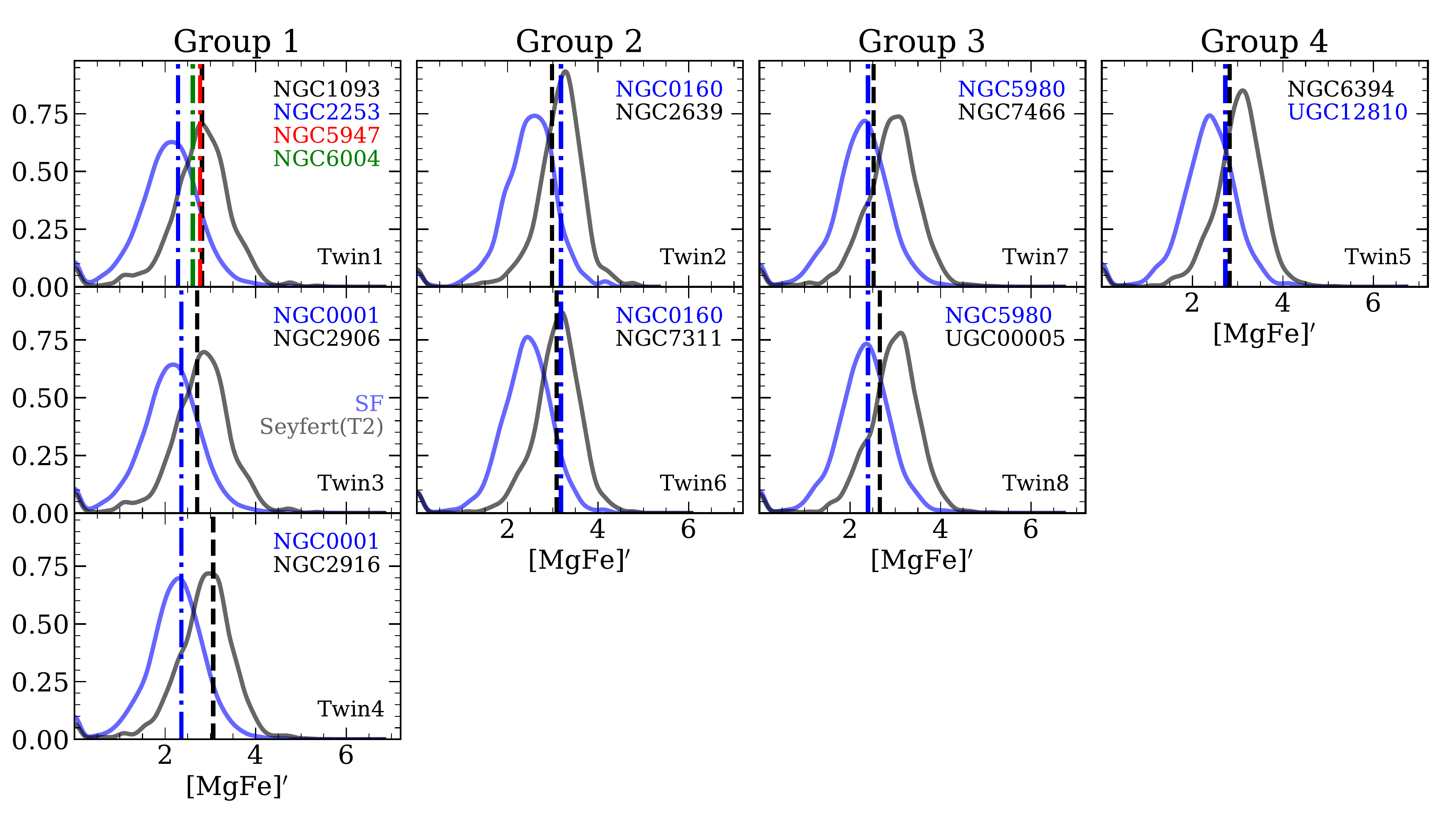}
    \caption{Same as Fig.~\ref{fig:D4k_SDSS} for the 
    distribution of [MgFe]$^\prime$ line strengths.}
    \label{fig:MgFe_SDSS}
\end{figure*}

\subsubsection{Age Sensitive Indices}
\label{sec:age_LSdiff_apertures}

\begin{figure*}
    \centering
    \includegraphics[width=\linewidth]{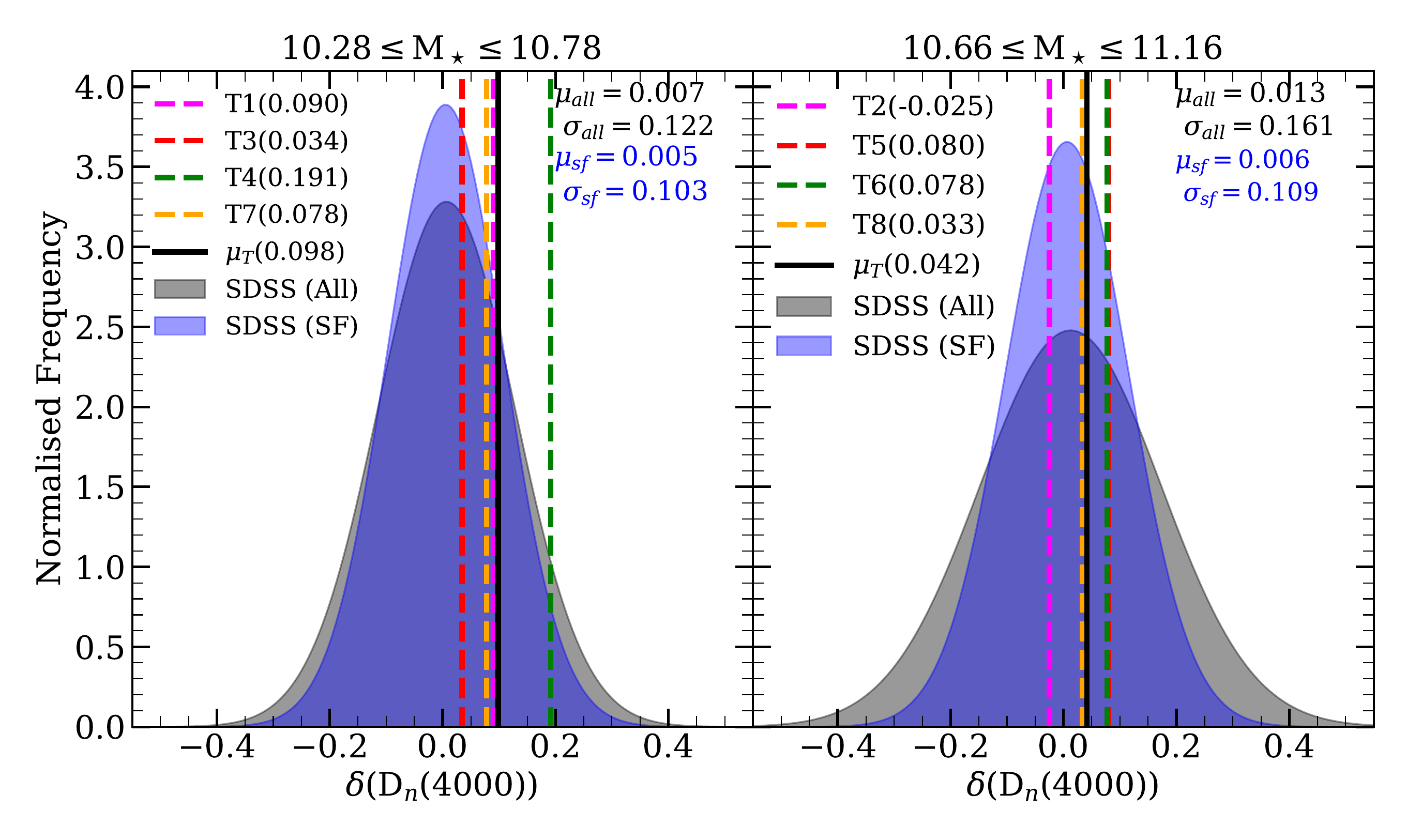}
    \caption{The significance of the results is tested with the
    relative difference of the D$_n$(4000) index between galaxies in
    each twin pair, contrasted with random samples from SDSS (assuming
    Gaussian distributions).  The blue distribution corresponds to random
    pairs of star-forming galaxies in SDSS.  The grey distribution
    corresponds to random sets where both star-forming and type~2 Seyfert AGN
    from SDSS are considered. The panels are split with respect to the
    stellar mass,
    $10.28 \lesssim \log\, \mathrm{M_\star/M_\odot} \lesssim 10.78$
    (left) and
    $10.66 \lesssim \log\, \mathrm{M_\star/M_\odot} \lesssim 11.16$
    (right).  The vertical lines mark the observed relative differences
    in the twin pairs, as labelled.}
    \label{fig:RD_D4k}
\end{figure*}

\begin{figure}
    \centering
    \includegraphics[width=\linewidth]{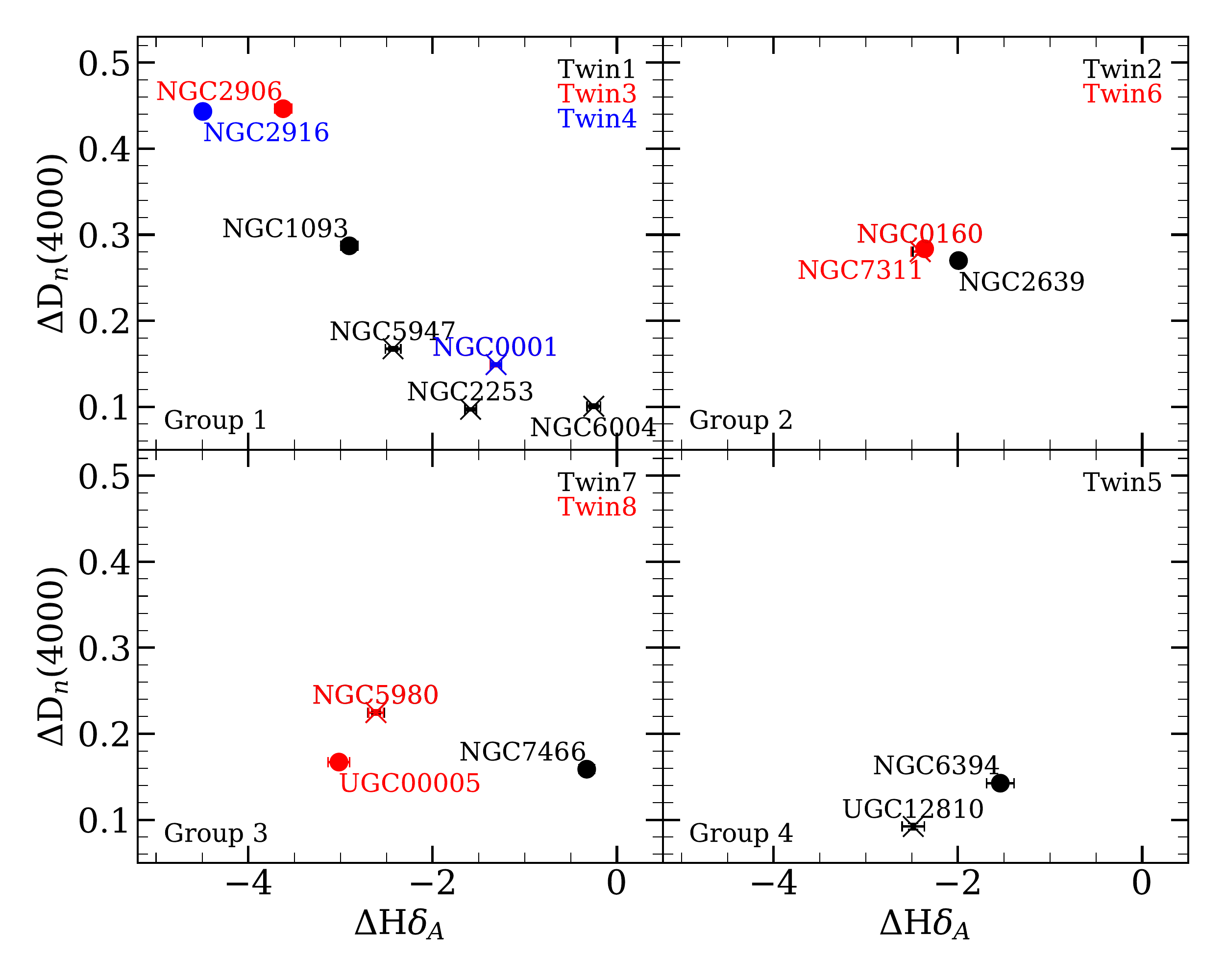}
    \caption{Line strength difference between the central region (1.5\arcsec)
    and the largest aperture (1.5\,R$_{\rm eff}$), regarding H$\delta_A$ and 
    D$_n$(4000). The circular data points denote galaxies 
    hosting an AGN, whereas the cross symbol shows the SF counterparts.
    Each colour identifies twins, as labelled in each panel. Note that 
    some twins have the same SF galaxy (NGC0160, NGC0001 and 
    NGC5980), so those appear repeated in the figure, hence they 
    have overlapping data points. The uncertainties are
    produced with a set of Monte Carlo realisations.}
    \label{fig:GDiff_Age}
\end{figure}

Fig.~\ref{fig:GDiff_Age} shows $\Delta$D$_n$(4000) vs
$\Delta$H$\delta_A$ for galaxies split in panels with respect to the four
groups. The different colours denote different twins, as labelled.
AGN galaxies are indicated by a circular data point, whereas SF
galaxies are shown by a cross. 
G1 galaxies -- i.e.  twins 1, 3 and 4 in the top-left panel, 
show a clean
trend where galaxies hosting the AGN have a greater difference between
the central and outer regions in both indices.  AGN galaxies have a
difference of $\Delta$D$_n$(4000)$\gtrsim 0.2$ and
$\Delta$H$\delta_A \lesssim - 2.6$.
Both AGN and SF galaxies feature an older central part compared to the 
outer regions, favouring the idea of inside-out quenching \citep[e.g.][]{Spin:2018, Kelvin:18}.
In G2, twins 3 and 4, the 1.5\,$R_{\rm eff}$  aperture spectra show 
similar 4000\AA\ break, regardless of galaxy type (see Fig.~\ref{fig:Bimod_Min}).
In contrast, 
the SF systems have a consistently 
lower difference between central and outer apertures 
$\Delta$D$_n(4000)\lesssim$0.2 and
$\Delta$H$\delta_A \gtrsim -2.6$, indicating that the presence of 
an AGN preferentially quenches the central region. 
The G4 twins (bottom-right panel) show a similar trend, where the radial difference 
in 4000\AA\ break strength is greater in the AGN system, although the difference
here is more subtle. Note that G4 is defined by a twin where
the AGN is in the ``lower'' part of the GV and the SF is in the BC.
This figure shows that, in addition, the radial trends
are shallower, and less distinguishable between AGN and SF galaxies.

G2 galaxies show substantial gradients, once more suggesting older
populations in the central regions. However, there is no clear
difference between AGN and SF members, a result that could be expected
from the fact that G2 twins {\sl both} have the central spectra in the
RS, and the 1.5\,R$_{\rm eff}$ spectra in the upper portion of the BC.
We can thus assume that the SF galaxy, while
being classified as SF, is a system closer to end of its star formation
cycle. This is further supported by their earlier disc morphological
classification (Sa). In a general context, group G2 
is an anomaly in our sample. In G3 both AGN and 
SF galaxies reside in the lower part of the GV. Here, a substantial
difference is found in the radial gradient of 4000\AA\ break strength, with
larger variations in the SF systems with respect to the AGN,
whereas H$\delta_A$ has similar variations within this
group. 

\subsubsection{Metallicity Sensitive Indices}
\label{sec:Met_params_apertures}
\begin{figure}
    \centering
    \includegraphics[width=\linewidth]{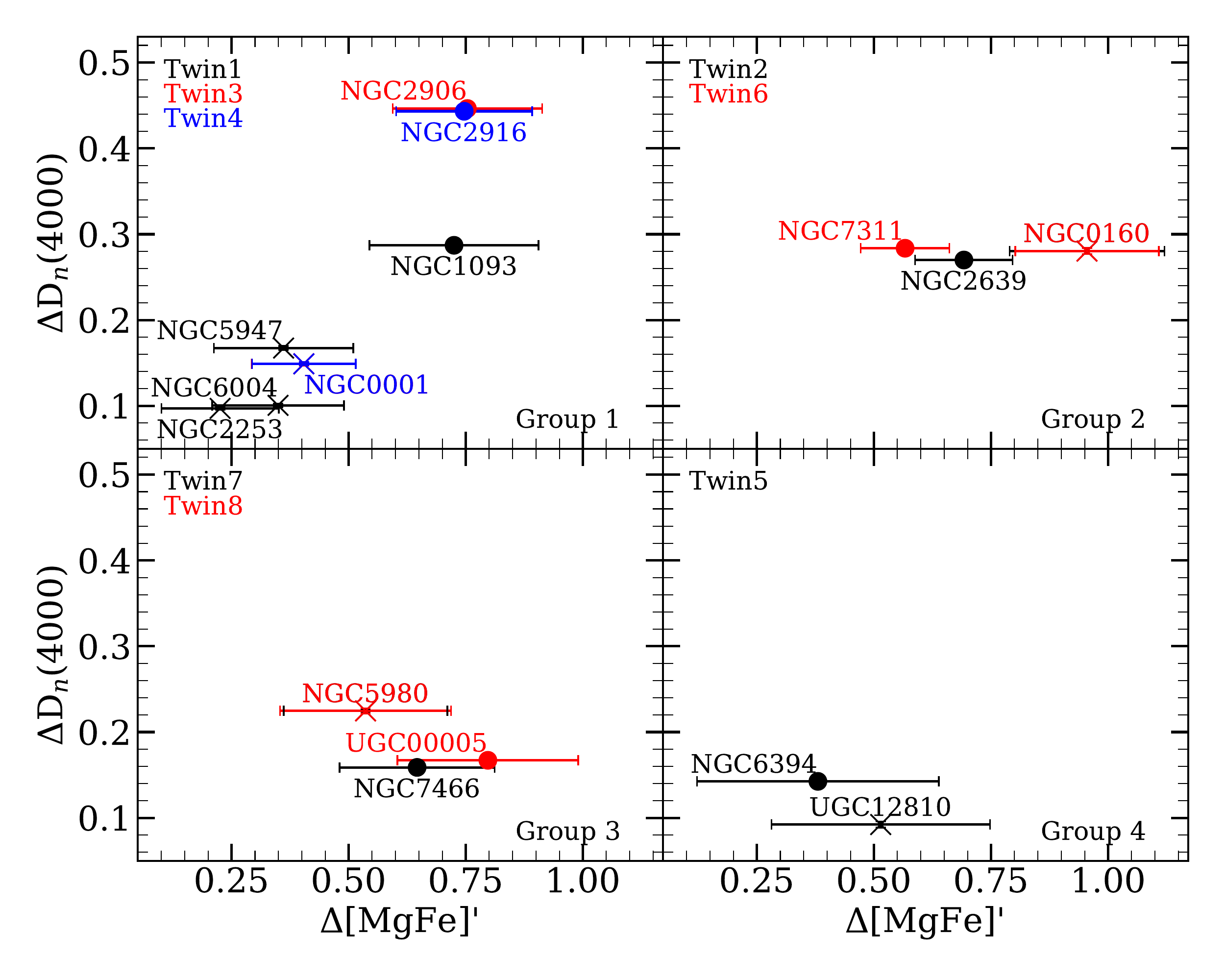}
    \caption{Same as Fig.~\ref{fig:GDiff_Age}, showing a line strength
    diagram comprising D$_n$(4000) vs [MgFe]$^\prime$.}
    \label{fig:GDiff_Met}
\end{figure}

Fig.~\ref{fig:GDiff_Met} shows the equivalent of Fig.~\ref{fig:GDiff_Age},
replacing H$\delta_A$ with the metallicity-sensitive index 
[MgFe]$^\prime$, following the same labelling and colour coding.
Once more, G1 galaxies show the clearest trend, where AGN systems have a greater
difference in [MgFe]$^\prime$ ($\Delta$[MgFe]$^\prime\gtrsim 0.6$),
with respect to their SF twin counterpart
($\Delta$[MgFe]$^\prime\lesssim 0.5$).  
This trend suggests 
a more metal rich population in the central regions 
regardless of galaxy type, as $\Delta$[MgFe]$^\prime\gtrsim$0,
however, the AGN consistently have higher gradients.
G3 and
G4 galaxies show no clear difference between AGN and SF galaxies,
given the error bars, but they consistently feature negative 
radial gradients in metallicity ($\Delta$[MgFe]$^\prime\gtrsim 0.25$).
A slightly more significant trend is shown
in G2, where in both twins, SF galaxies have a greater difference
between apertures, $\Delta$[MgFe]$^{\prime}\gtrsim$0.8, compared to
their AGN counterpart $\Delta$[MgFe]$^{\prime}\lesssim 0.8$. Note, there are
large uncertainties associated to the data points.

These results suggest an alternate view of the radial extent of AGN
feedback. If we were to believe low-to-intermediate AGN 
activity only affects the
formation history within a relatively small region around the centre,
we should consistently find greater differences in the AGN galaxy of
each twin. The data are not conclusive, and reveals mixed
distributions. Furthermore, del Moral-Castro et al. (in prep.) will 
investigate these results further using spectral fitting.
The implications of this are discussed in
Sec.~\ref{sec:Disc}.

\subsection{SSP Parameters}
\label{sec:SSP_params_fid}
In this section, the SSP equivalent ages and metallicites are 
estimated for the twin galaxies within the 3 different 
apertures. We make use of the MIUSCAT population synthesis models \citep{Vaz:2012}, 
constructing a grid consisting 
of 8192 synthetic spectra, 128 ages varying from 0.1 to 13.5\,Gyr
in a logarithmic scale. Similarly, the metallicity 
varies from [Z/H]=-2.0 to +0.2 also in log steps. 
The best fit to the SSPs adopts a $\chi^2$ statistic defined
as follows:
\begin{equation}
  \chi^2(t,Z) = \Sigma_i\left[ \frac{\Delta_i(t,Z)}{\sigma_i}\right]^2, 
\label{eq:chival}
\end{equation}
where $\Delta_i(t,Z) = O_i - M_i(t,Z) - \delta_i$ is the difference
between the observed and model index with an offset ($\delta_i$) for
the $i$th index.  This offset is introduced due to the high S/N in our
observed spectra, at a level where the SSP models are not capable of
fully reproducing all the details. The uncertainties $\sigma_i
= \sqrt{\sigma_{i,err}^2 + (0.05 O_i)^2}$, encapsulate both the
statistical error in the index along with an extra term amounting to
$5\%$ of the index value.  This way we account for both the
systematics associated with our methodology and include a conservative
uncertainty in our calculations. The $\chi^2$ statistic involves a set
of seven different spectral indices: D$_n$(4000), H$\delta_A$ and
H$\gamma_A$ (age sensitive indices), and Mgb, Fe5270, Fe5335 and
[MgFe]$^\prime$ (metallicity sensitive indices). The likelihood
corresponding to the SSP-equivalent estimates of age and metallicity
are obtained by marginalising over the unwanted parameter. While some
indices are described as ``age-sensitive'' and others are
``metallicity-sensitive'', all of the indices suffer from the
age-metallicity degeneracy \citep{Worthey:94}.

\subsubsection{Age parameters}
\label{sec:age_params_SSP}
\begin{figure*}
    \centering
    \includegraphics[width=55mm, height=55mm]{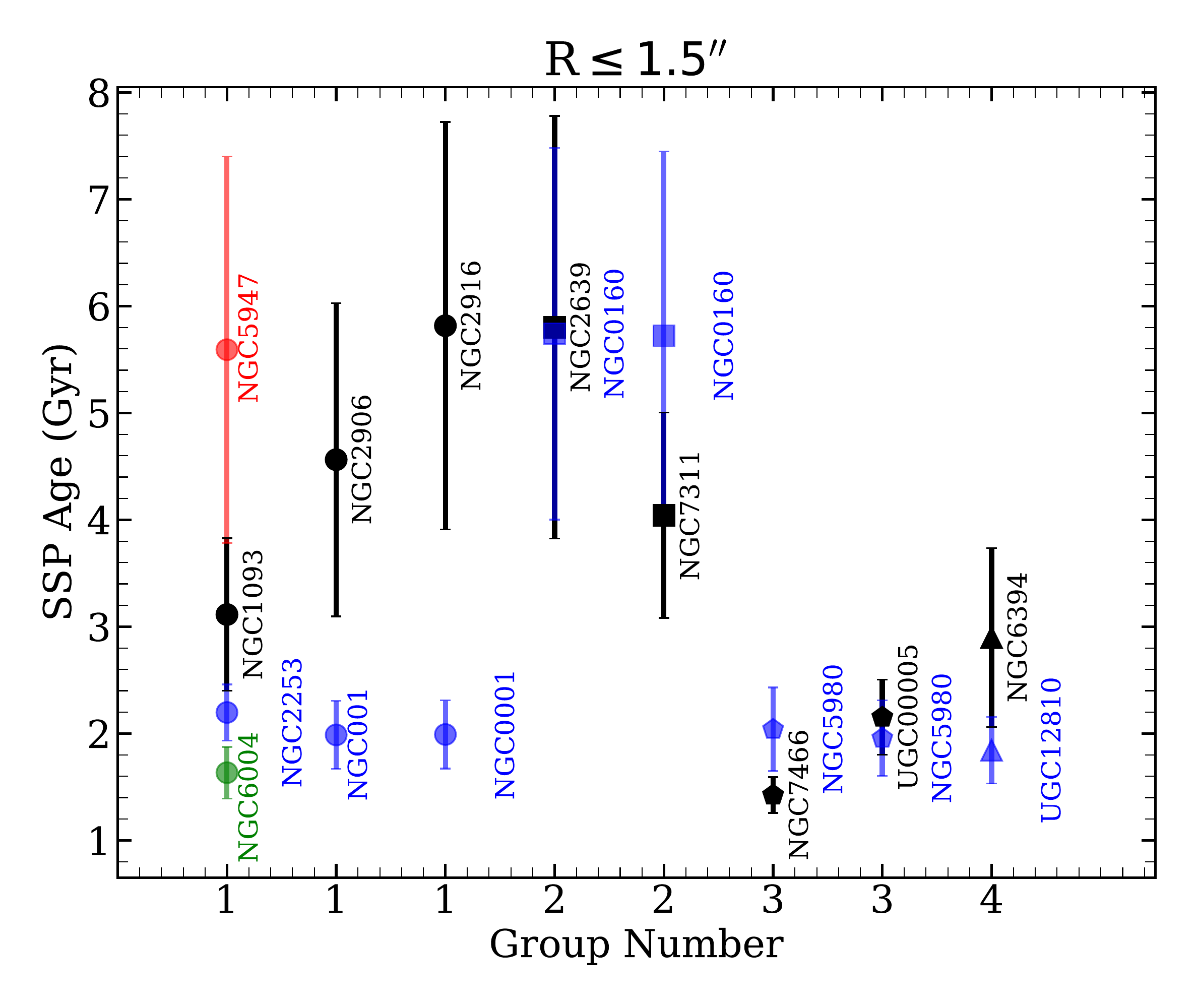}
    \includegraphics[width=55mm, height=55mm]{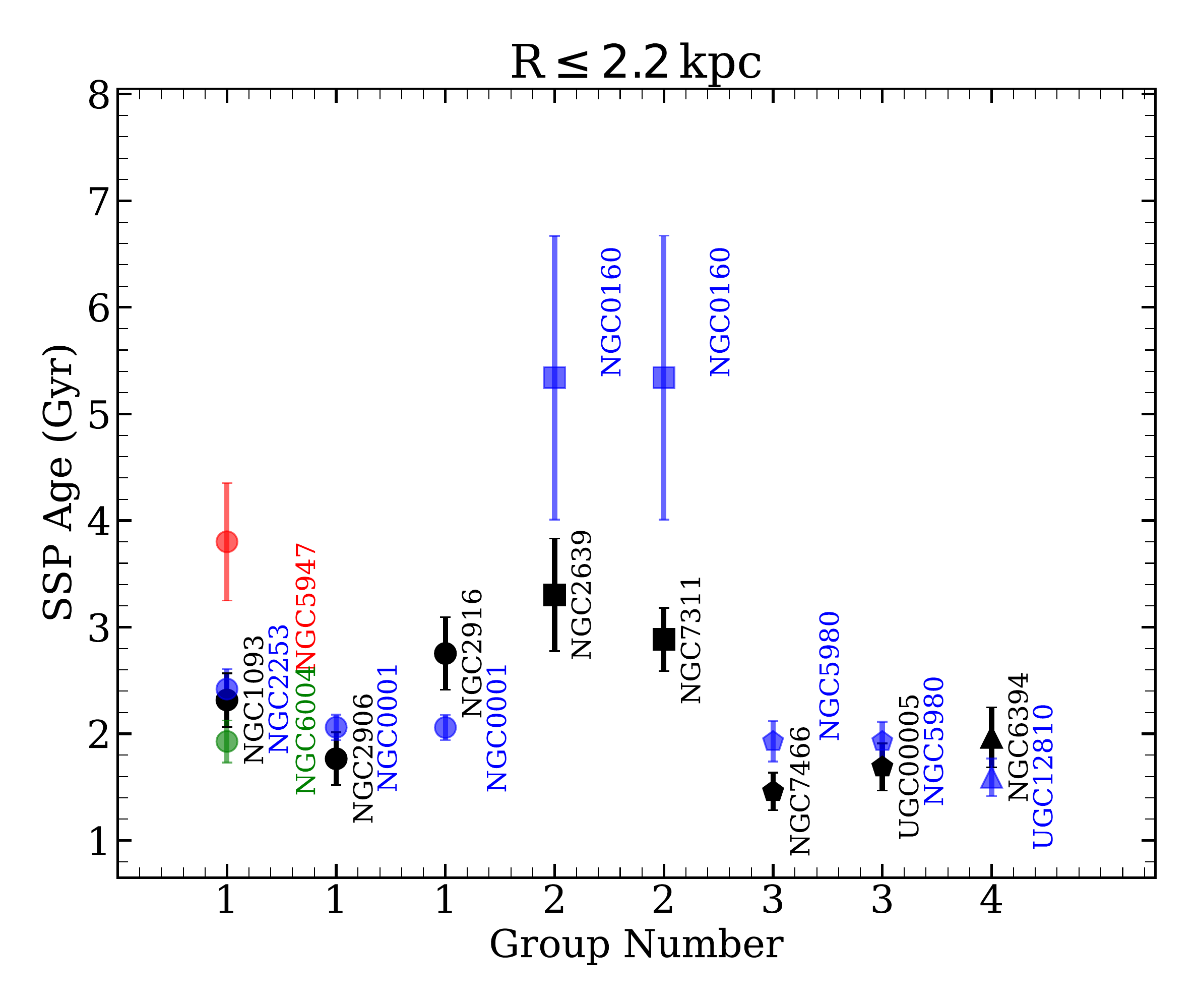}
    \includegraphics[width=55mm, height=55mm]{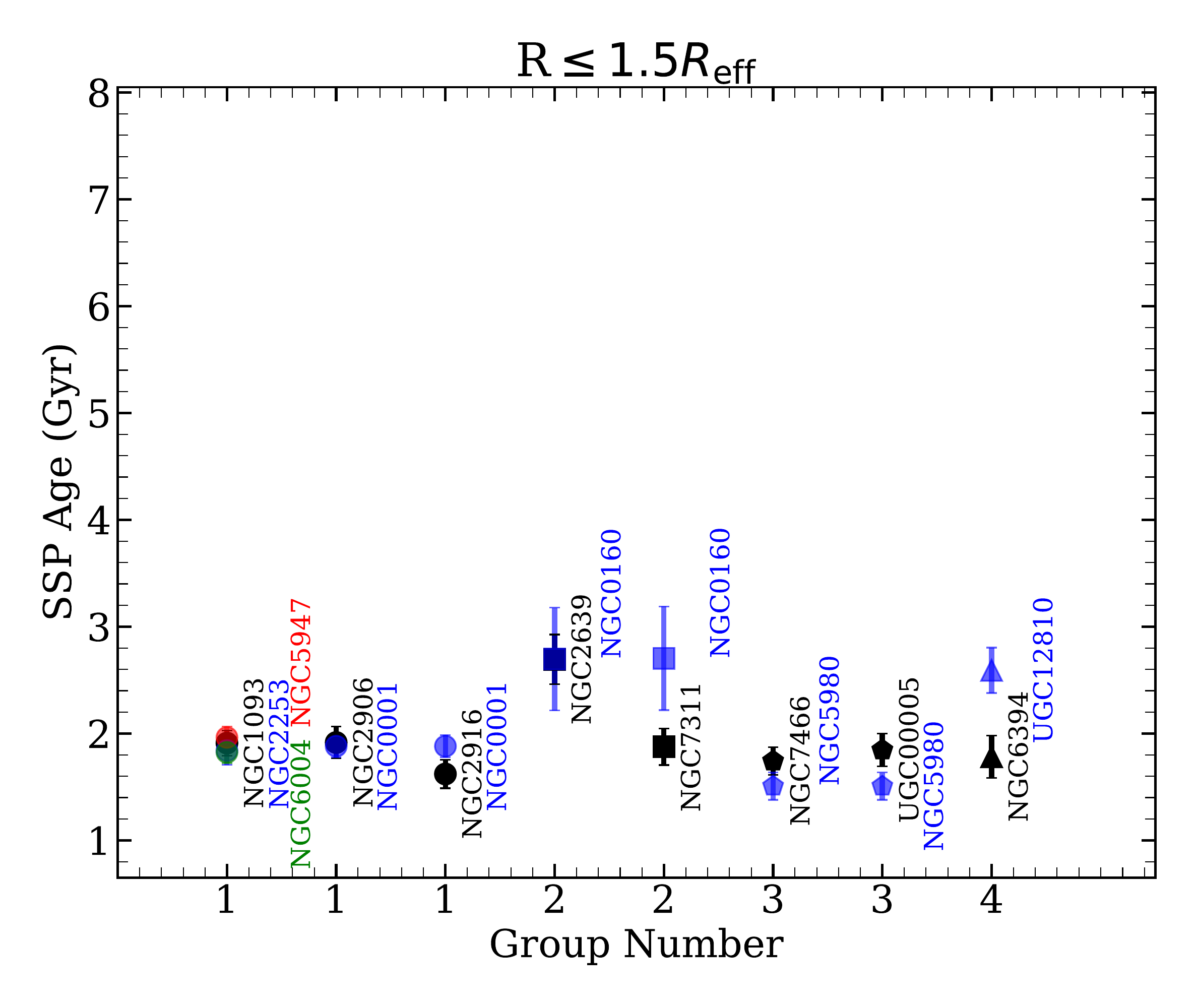}
    \caption{SSP equivalent average ages for twin galaxies in different
    groups. The black and coloured data points, blue, green and red,
    show the results for AGN and non-AGN galaxies, respectively.
    The error bars are given at the 1$\sigma$ confidence level, 
    obtained by calculating the normalised likelihood value, which 
    makes use of the different $\chi^2$ values. The error on each of 
    the individual line strengths are obtained by making 100
    Monte Carlo realisations of the spectra - which is used to 
    calculate the $\chi^2$ with best fit. From left to right, we show the
    results for
    twins 1, 3, 4, 2, 6, 7, 8, and 5. Furthermore, the  circle, square,
    pentagon and triangle symbols identify groups 1, 2, 3
    and 4, respectively.}
    \label{fig:Fid_SSP_Age}
\end{figure*}

Fig.~\ref{fig:Fid_SSP_Age} shows the SSP equivalent age for different
groups within different apertures. The left, middle and right panels show 
the ages for the central 1.5\arcsec, 2.2\,kpc and 1.5\,R$_{\rm eff}$ apertures, 
respectively. The black data points show the ages of the AGN galaxy in
each twin, while the other colours represent the ages of SF
galaxies. Circle, square, pentagon and triangle symbols denote G1, G2, G3 and
G4, respectively. We find $80\%$, $40\%$ and $60\%$ of the AGN galaxies
in the twin being equal or older than their SF counterpart for G1 galaxies inside
of the 1.5\arcsec, 2.2\,kpc and 1.5\,R$_{\rm eff}$ apertures, respectively.
In G2, AGN galaxies have equal or younger average ages compared to their SF
counterpart 83$\%$ of the time, even though in the evolutionary
sequence, within a 1.5\arcsec aperture, we find these AGN to have a 
stronger 4000\AA\ break. G3 twins show no clear trend in the inner regions (1.5\arcsec
and 2.2\,kpc) but show older populations in AGN with respect to SF galaxies, for
both twins, in the 1.5\,R$_{\rm eff}$ aperture.  G4 follows the
opposite trend, where the AGN galaxy is older within both 1.5\arcsec and 2.2\,kpc
apertures, but is younger, than the SF galaxy, when considering the
largest aperture.

If we consider all the twins in the different groups, in all
apertures, we find that 50\% of twins show equal or older populations in the
AGN galaxies ($70\%$ in 1.5\arcsec, $30\%$ within 2.2$\,$kpc and
$50\%$ for the largest aperture). This indicates that the AGN exerts
the most impact within the most central region of the galaxy.

\subsubsection{Metallicity parameters}
\label{sec:Met_params_SSP}
\begin{figure*}
    \centering
    \includegraphics[width=55mm, height=55mm]{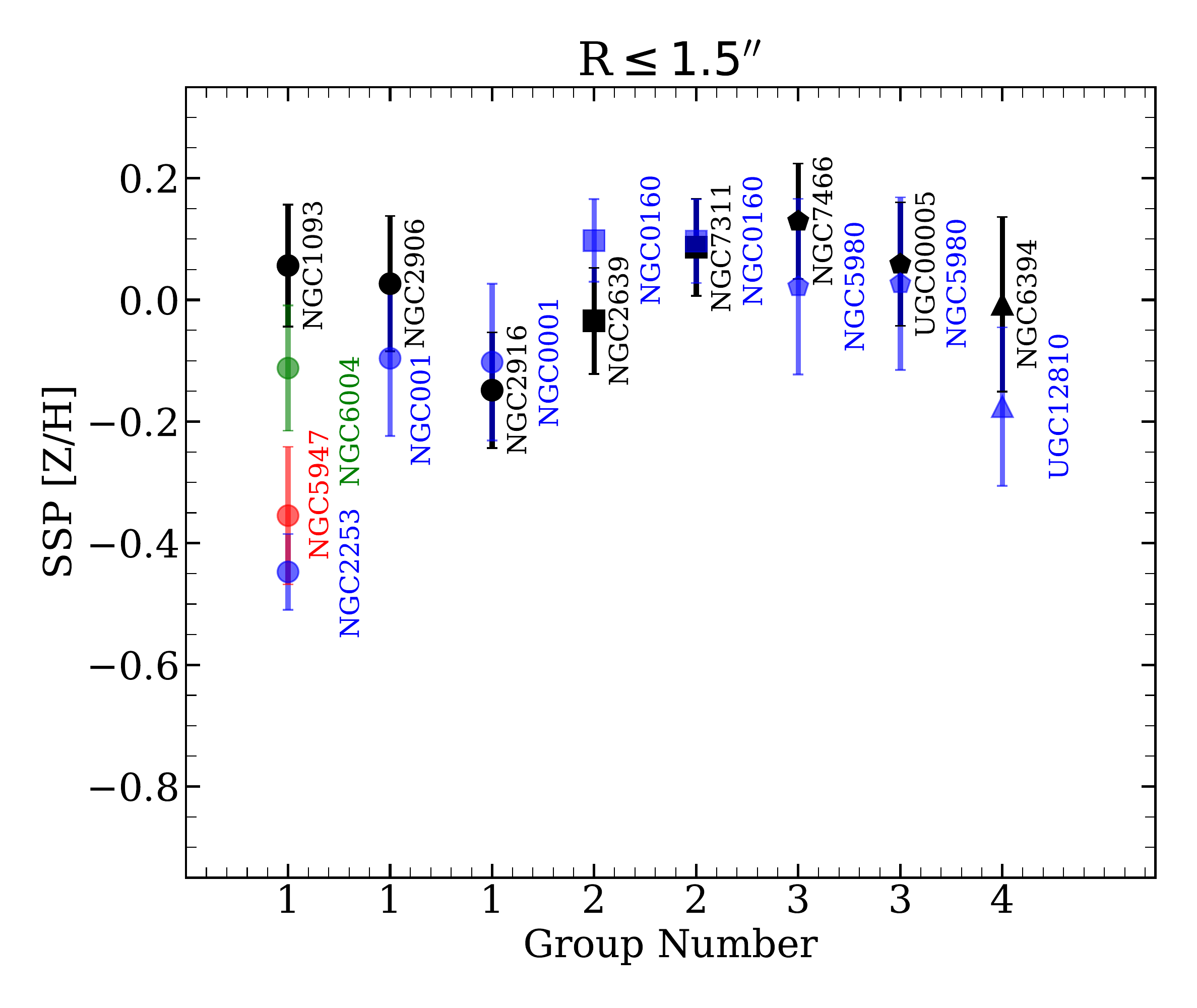}
    \includegraphics[width=55mm, height=55mm]{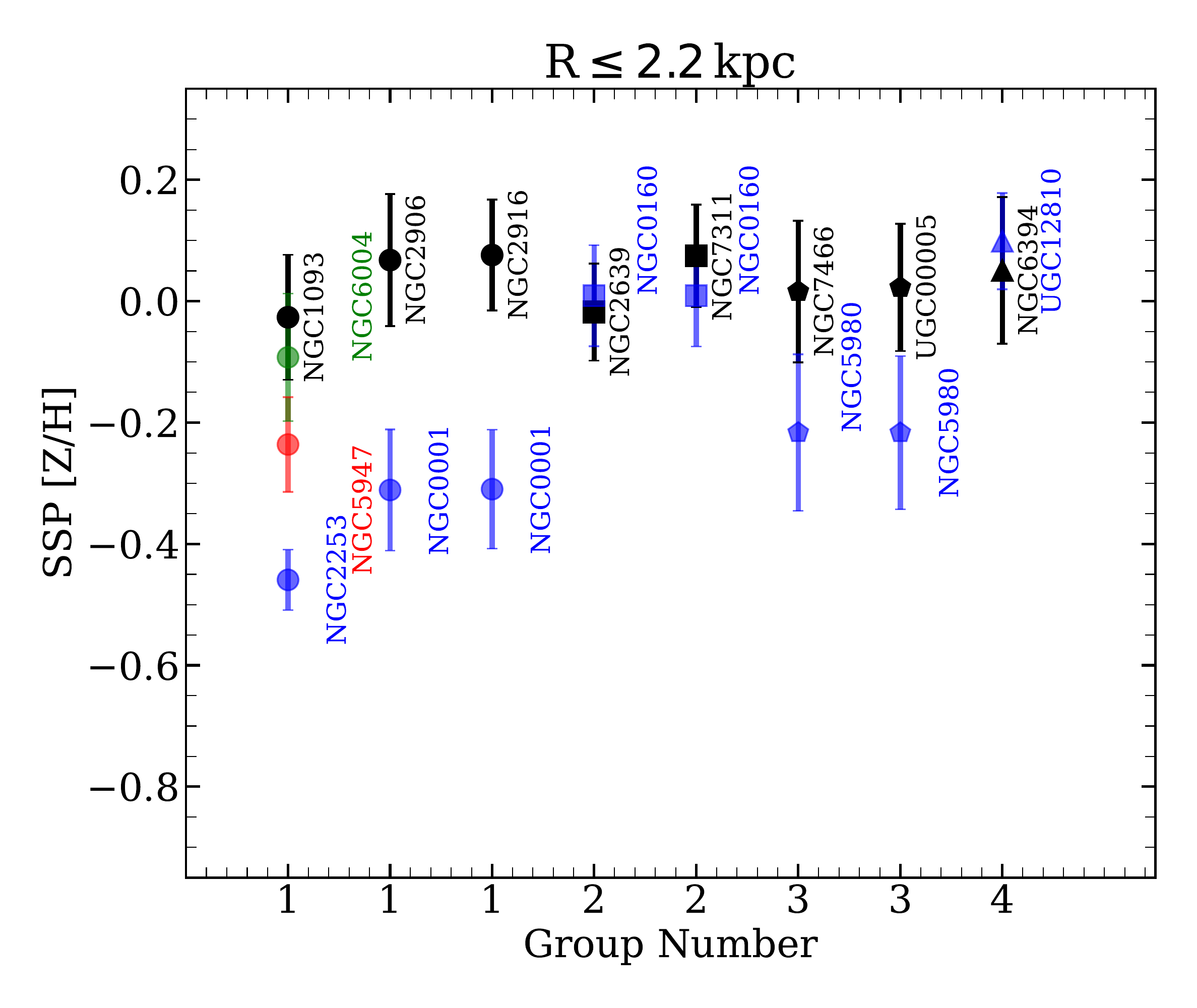}
    \includegraphics[width=55mm, height=55mm]{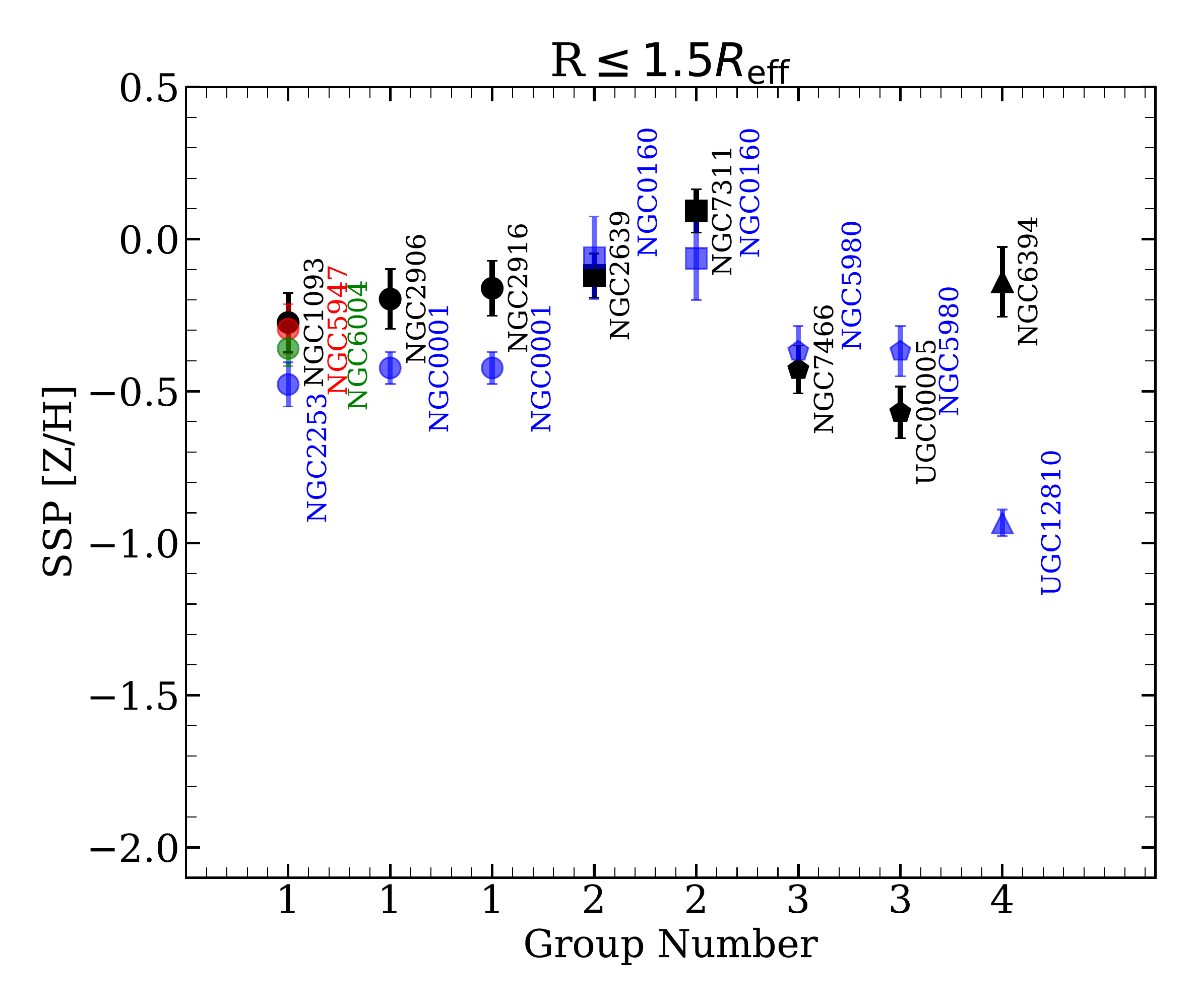}
    \caption{Same as Fig.~\ref{fig:Fid_SSP_Age} for the 
    SSP equivalent metallicity of the twin pairs, following an identical notation.}
    \label{fig:Fid_SSP_Met}
\end{figure*}

Fig.~\ref{fig:Fid_SSP_Met} shows the SSP-equivalent metallicities,
from left to right, within the central 1.5\arcsec, 2.2\,kpc, and
1.5\,R$_{\rm eff}$ apertures, respectively. 
The symbols and colours are equivalent to those
shown in Fig.~\ref{fig:Fid_SSP_Age}.
Similarly to the age estimates,
within 1.5\arcsec,
G1 galaxies have AGN that are more metal rich than the SF galaxies 80$\%$ of 
the time. G2 and G3 show no clear trend. 
Finally G4, also shows AGN galaxies to be 
more metal rich. In the 2.2$\,$kpc aperture 
G1 and G3 host AGN galaxies that are equal or more metal 
rich than their SF counterpart 100$\%$ of the times. 
G4 shows the SF galaxy to be more metal rich, however
the differences are within 1$\sigma$. 
For the more metal rich AGN galaxies, we sometimes find them to be
younger than their SF counterpart, perhaps a sign of the age-metallicity
degeneracy.   However in
Appendix~\ref{sec:AB_Deg}, we show the bivariate confidence levels in
age and metallicity for 2 twins, rejecting a substantial bias from 
this degeneracy, specifically for twins where we find the AGN galaxy 
to be more metal rich.
Therefore, at least within a 2.2\,kpc aperture, the metallicity
trend indicates that the presence of an AGN is potentially related to a 
different star formation history, with respect to the SF counterpart.
This is further backed
up by the spectra in the largest aperture, 1.5\,R$_{\rm eff}$ trends,
where we find that the AGN galaxies in G1 and G4 are more
metal rich compared to their SF counterpart. Note, while we find
strong evidence showing AGN galaxies are likely to be more 
metal rich, some twins show large overlap between AGN and SF 
twins.
Furthermore, rather than looking at an individual group within a fixed 
aperture, if we look at all the different groups and in all 
apertures, we find AGN galaxies to be more metal rich 73$\%$ 
of the time compared to their SF counterparts. Breaking this
into different apertures, we find 70$\%$, 80$\%$ and 70$\%$ of AGN galaxies 
to be more metal rich in the 1.5\arcsec, 2.2\,kpc, and 1.5\,R$_{\rm eff}$,
respectively. Note, some of these galaxies have values, which 
are compatible within 1$\sigma$.

\section{Discussion}
\label{sec:Disc}
This paper explores the transitional role of AGN activity by analysing
pairs of galaxies defined as twins, inasmuch as they feature similar
large galaxy-scale properties but one presents AGN activity and the other one does
not. The definition of these twin pairs follows the methodology laid
out in \citet{Ig:2020}.  In this section we discuss the potential
implications of our results, from the radial influence of AGN activity
on the galaxy, to testing two alternative hypotheses, either an ``on-off''
AGN switching mode, or the
possibility of the twins having different evolutionary paths. Owing to
the small sample size (10 twins comprising 8 AGN and 7 SF galaxies due to
some AGN galaxies have multiple SF pairs),
we mostly find mixed trends in
the behaviour of the individual twins. A key reason for such
difference could be due to their differing location on the
evolutionary sequence, as galaxies on the BC have different stellar
populations with respect to those on the GV or RS. Nevertheless, a
statistical comparison of the CALIFA spectra measured within 2.2\,kpc
with respect to a more general sample from SDSS suggests that many of
these twins -- twin pair 1, 3, 4, 5, 7 and 8 -- are representative of
the larger sample, taken from SDSS  (see Figs.~\ref{fig:D4k_SDSS},
\ref{fig:HdA_SDSS}, and \ref{fig:MgFe_SDSS}), thus providing strong
motivation for similar studies in larger data sets.

\subsection{Potential Caveats}
\label{sec:Caveats}

This paper is meant to present a methodological approach that can be
used to assess the role of AGN in quenching star formation in
galaxies, focusing on the diagram that presents the transition Blue
Cloud/Green Valley/Red Sequence as a fundamental evolutionary sequence 
(Fig.~\ref{fig:Bimod_Min}). While larger data sets are needed to draw
strong conclusions, this paper illustrates how this type of
information can be retrieved from a reduced sample of twin pairs. We
discuss here some of the caveats one should be aware of.

Note for the selection of the AGN and SF galaxies, we use the BPT
classification scheme \citep{BPT:81}. The analysis is applied to the
central spaxel of each galaxy. This could introduce a systematic trend, 
as the separation between AGN and SF activity on the BPT diagram
mainly depends on the hardness of the radiation field, whereas other
parameters, such as metallicity, pressure, the ionisation parameter, 
or the presence of shocks, will also affect 
the emission lines adopted in this classification \citep{Kewley:19}. Furthermore,
previous studies  have found the SF region
of the BPT diagnostic diagram to be systematically metal poorer than
in the AGN region \citep{Kewley:2013, Ji:2020}.
In this work, the AGN classification is based on the 
central spaxel, with an effective resolution of 1\arcsec, whereas our
analysis of line strengths and SSP age and metallicity are carried out
in radii of 1.5\arcsec, 2.2\,kpc, and 1.5\,$R_{\rm eff}$.

Another caveat is the small sample size of our data; we have only 8
different AGN galaxies and 7 SF galaxies. Throughout our study we have
assumed that G1 and G4 galaxies are to be most representative of the
larger sample, as these galaxies have D$_n$(4000), H$\delta_A$ and
[MgFe]$^{\prime}$ values located at the peak of their respective SDSS
galaxy distribution. This suggests that if we were to carry out a
study on a larger sample, we would expect many of the twin pairings to
be similar to those belonging to G1 and G4.  However, this is
inconclusive due to our statistically small sample. While G2 and G3
twins have SF galaxies located away from the peak of the SDSS galaxy
distribution, a larger study might find G2 and G3 to be more
representative of ``twin'' samples. This could be owing to most SF
galaxies not being representative of the ``twin'' control sample of
AGN galaxies.  However, the key result from such grouping should be
their diversity even amongst ``twins'', indicating the effect of AGN
on the stellar population of their host to be of a complex
nature. Such diverse grouping motivates the need for a larger study.

\subsection{Radial population variation}
\label{sec:rad_trend}
Both AGN and SF galaxies present mostly older stellar populations  
in the inner part of the galaxy, with a decreasing gradient as we
move radially out (Sec.~\ref{sec:result1}),
in agreement with del Moral-Castro et al. (in prep.),
who adopt a full spectral fitting approach. 
\citet{San:2018} also find similar results in the MaNGA survey, 
using the derived star formation rate and gas density of their sample.
Similarly, studies 
of stellar populations at different radii in CALIFA 
unveil similar trends, noting a decrease in age and
metallicity for different types of galaxies at increasing radii
\citep{Bitsakis:2019,Lacerna:2020,Kalinova:2021}. 
All three results favour inside-out quenching 
\citep{Lipari:1994,Tacchella:2015,Li:2015,Iris:20},
which suggests a galaxy may run out 
of gas without any interaction with other galaxies, involving
internal processes, for instance, through secular evolution.

\subsection{``On-off'' AGN hypothesis}
\label{sec:onoff_AGN}
The next two subsections propose the interpretation of the results with
two alternative scenarios. While the sample size in this work is rather
small, preventing us from producing strong conclusions, these simple
scenarios can be adopted to larger samples, to assess the role of AGN
in quenching star formation.

\begin{figure}
    \centering
    \includegraphics[width=\linewidth, height=80mm]{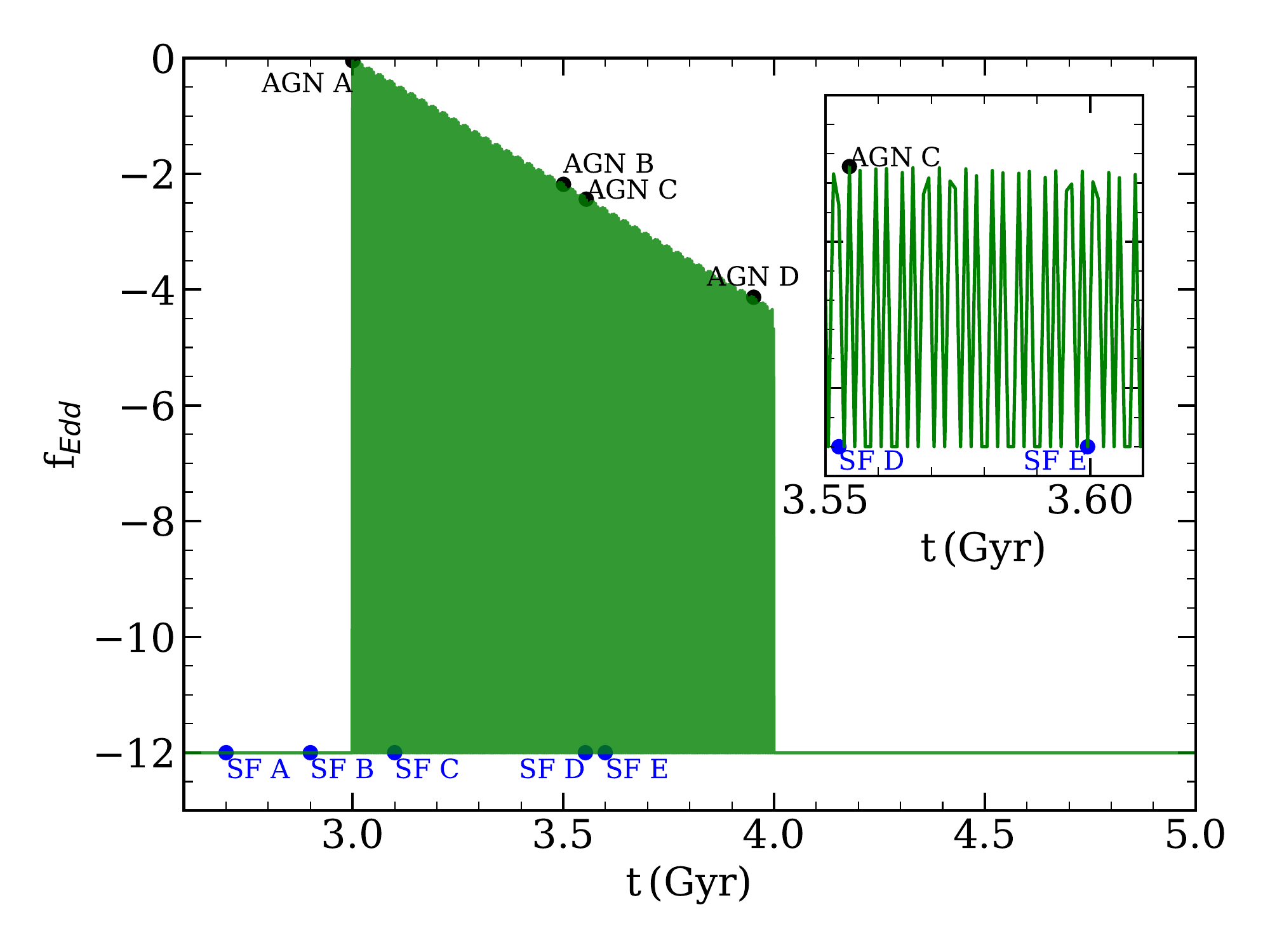}
    \caption{The fluctuating activity of the AGN in a galaxy is
    illustrated by this simple schematic that plots the evolution of the Eddington 
    ratio of the SMBH (green line). The blue and black 
    dots represent the typical Eddington ratio (f$_{Edd}$)
    of galaxies identified as SF 
    and AGN, respectively. The labels SF A, B, C , D and E (AGN A, B, 
    C and D) indicate
    different ages, i.e. different stages of evolution. The inset
    panel zooms in on a specific time frame to show the oscillatory 
    behaviour of the AGN.}
    \label{fig:Edd_Aging}
\end{figure}

The first one invokes a simple ``on-off'' AGN mode to explain the results.
This scenario is motivated by the fact that most galaxies experience
AGN events but some happen to be in an active phase (``on'' state),
while others are dormant (``off'' state).  The timescale of individual
``on-off'' events is expected to be around $\sim 10^{5}\,$yr, where
the whole life cycle lasts for $\sim 10^{7} -
10^{9}$\,yr \citep[see, e.g.,][]{Schawinski:15}.  In such a cyclic behaviour, it is
possible to explain the similarities and differences in twin pairs.
Fig.~\ref{fig:Edd_Aging} illustrates this behaviour, with a 
schematic diagram of the 
variation in the Eddington ratio of the SMBH (f$_{Edd}$), in green,
throughout a galaxy's life cycle.
The dots give examples of galaxies that will be 
identified as AGN (black) or SF (blue) when classified on the BPT diagram, respectively. 
In the ``on-off'' model, the behaviour in f$_{Edd}$ will rapidly quench star formation when the AGN is in an ``on'' mode (high f$_{Edd}$),
at least within the central region, resuming the star formation activity once it goes back to the 
``off'' mode (low f$_{Edd}$) - depending on whether enough  
gas available for star formation is left over. This ``on-off'' behaviour operates over
timescales that are much shorter than the sensitivity of line strengths, and
even emission lines (from the ionising photons of massive stars). Therefore, 
we should see a gradual decrease in the SFR resulting in an ``ageing'' behaviour 
(a gradual decrease in star formation over a more extended time interval), 
rather than a sharp ``quenching'' \citep{Casado:2015, Caballero:2021}, i.e. a 
sudden stop in star formation. 
Note, this is especially true as all our galaxies have late-type morphology. 
Therefore, owing to the slow ``ageing'' of galaxies, AGN and SF in the sample 
``twin'' pairing (AGN B or C with SF D or E) should have line strengths that are
indistinguishable between  them if they had similar formation time, 
especially for D$_n$(4000) and
[MgFe]$^{\prime}$ within this hypothesis. We emphasize that differences could
also be due to a substantial difference in the evolutionary stage, for instance,
pairing AGN B or C with SF A or B, in Fig.~\ref{fig:Edd_Aging}.

G1 features the largest variation 
between AGN and SF systems, where AGN galaxies, measured within a 1.5\arcsec
aperture, reside near the RS (AGN D) but SF galaxies reside either in 
the BC or GV. 
In addition, G1 twins display a greater radial difference between the 
central region and the largest aperture, 1.5\,R$_{\rm eff}$, for AGN galaxies,
with respect to the SF counterparts
(see Fig.~\ref{fig:GDiff_Age}
and Fig.~\ref{fig:GDiff_Met}). Fig.~\ref{fig:Fid_SSP_Age} also 
shows generally an older population in the central region for
AGN galaxies with respect to their SF counterpart. Such behaviour
seems antithetical to the ``on-off'' AGN mode, however it 
can still be applicable, provided the AGN twin  has 
a different formation time compared to the SF equivalent. This could
hint towards the AGN galaxy being formed earlier than their SF twin
counterpart, as our selection criteria does not match for formation
time. Therefore, 
these AGN galaxies may be in a state closer to the end of their 
AGN life cycle, whereas the SF galaxies -- while still experiencing
some form of AGN activity -- have been formed in more recent cosmological times.
The G4 twin further supports 
this idea, as the AGN is located close to the GV but their
SF galaxy is located in the BC. However, since the AGN galaxy can be thought of 
as just entering the GV, we can assume it lies in a state more
representative of the middle of its AGN life-cycle (AGN B or C), 
resulting in similar 
properties between AGN and SF counterparts -- unlike the G1 twins
(see Fig.~\ref{fig:GDiff_Age} and ~\ref{fig:GDiff_Met}). Finally, 
galaxies in both G1 and G4 have bars, which is thought to play an important role 
with conjunction to AGN to quenching of star formation \citep[see, e.g.,][]{Ellison:11,PSB:11}.
However, the interpretation of such results is beyond the scope of
this paper.

G2 and G3 twins have AGN and SF galaxies with more similar 
population properties, regardless
of the aperture size. There is 
no clear distinction between AGN and SF galaxies at the
central region or in the largest aperture (see Fig.~\ref{fig:GDiff_Age}
and ~\ref{fig:GDiff_Met}), favouring the ``on-off'' AGN hypothesis, where
both AGN and SF galaxies have similar formation time (as in AGN B or C
paired with SF D or E in Fig.~\ref{fig:Edd_Aging}). 
G2 and G3 twins show similar behaviour between their AGN
and SF galaxies, however this could be due to different
reasons. G2 twins are both old, have very weak 
H$\alpha$ emission, and are classified as Sa, thus
featuring prominent central bulges. 
These properties indicate that the AGN in G2 may be undergoing 
maintenance mode activity \citep{Bari:19}, 
explaining the similarity found in the stellar 
populations of AGN and SF twin pairs. In contrast, 
G3 twins have strong H$\alpha$ emission lines for both AGN 
and SF members and they are classified as late spirals. Therefore
rather than invoking maintenance mode, we can think of these 
galaxies as undergoing more active ``on-off'' events at the 
start or middle of the AGN life cycle. Additionally, 
the analysis of the SSP equivalent ages (Fig.~\ref{fig:Fid_SSP_Age}) 
gives no clear difference between AGN and SF galaxies. 
Note that  previous studies, 
such as \cite{Lac:2020} and A19, noted
AGN hosts are more evolved than SF galaxies, as 
the former are mainly located in the GV, whereas the latter mostly reside
in the BC. However, in the comparison between this targeted set of twin galaxies, 
this behaviour
seems less clear. Within the ``on-off'' AGN hypothesis, this contrast 
should be due to their different formation times.

\subsection{``no-AGN'' hypothesis}
\label{sec:no_AGN}
At low and intermediate stellar mass 
(M$_\star\lesssim 10^{10.3}\,$M$_\odot$) there is ample evidence suggesting that 
stellar feedback and environmental mechanisms are sufficient 
to quench star formation  \citep{Feld:10, Naab:2014}. However,
at the massive end, AGN activity has been necessary to explain quiescent galaxies
\citep{JR:1998, Nel:18}. We propose an alternative simple model,
which is an alternative to the ``on-off''
model, to explain the properties of the twin pairs. In the ``no-AGN''
hypothesis we assume that,
within same stellar mass, some galaxies will quench their star
formation through AGN activity while others will (i) not experience
any AGN (ii) will not quench their SF through AGN but through
different physical mechanisms such as stellar feedback or 
morphological quenching mechanisms.
This scenario would imply that
in twin galaxies, we should find statistically significant differences
between the stellar populations of AGN and SF twin pairs.  Previous
work in the literature, such as \cite{Ig:2019} and \cite{Ig:2020},
have found strong evidence supporting the ``no-AGN'' hypothesis, where
they find AGN galaxies to consistently display higher angular momentum
than their SF counterpart. Note that variations in angular momentum
are related to larger timescales, that cannot be explained in the
context of a simple ``on-off'' AGN switch model (assuming similar
formation time for these ``twin'' systems), as the AGN duty cycle is
rather short-lived, of order $\sim 10^{5}\,$yr. Similarly,
these short timescales are not sufficient enough to explain 
the line strength differences we observe, as stellar population 
indicators vary over timescales of order $\sim$100\,Myr.

In this work, the analysis of G1 twins
generally shows older populations in AGN galaxies (Fig.~\ref{fig:Fid_SSP_Age}).
This trend, while possible, would be unlikely, if we were to
assume a simple ``on-off'' scenario. 
The estimation of SSP
metallicity, Fig.~\ref{fig:Fid_SSP_Met}, shows AGN galaxies are 
generally more metal rich compared to their SF counterparts, within
1.5\arcsec and 2.2\,kpc apertures.
Once more, such variation in metallicity is expected over 
longer timescales ($\gg 100\,$Myr), indicating a different chemical enrichment 
history, which implies timescales that are longer than those expected 
in the ``on-off'' AGN switching mode.
The analysis of the G4 twin yields a similar
result, where we find the metallicity of AGN galaxies consistently 
greater than their SF counterpart.

We emphasize the diversity found among galaxies in such a small sample,
reflecting the subtle role of AGN quenching over the timescales that can
be probed with stellar population studies. While we find evidence supporting
both models presented here, a larger set of twin pairs is needed
to assess the validity of the two alternative hypotheses.

\section{Summary}
\label{Sec:Conclusion}
In this paper, we have investigated the stellar
population properties of a carefully defined state-of-the-art sample of
twin galaxies \citep{Ig:2020}, selected from the 
CALIFA IFU survey. Galaxies in twins are expected to appear
undistinguishable from the point of view of size, mass,
morphology and inclination, with the only difference being
the presence or absence of an AGN. We project this sample onto a dust
resilient evolutionary plane spanned by D$_n$(4000) vs stellar mass;
following the methodology outlined in A19. The original
11 twin sample (20 twin pairs) is reduced to 8 twins (10 twin pairs), 
due to a new selection criteria
restricting differences in velocity dispersion and stellar mass,
to maximise the similarities between twins concerning the
stellar populations. We study stacked spectra 
within three different apertures -- the most central region,
(R$\leq$1.5\arcsec), a region that matches, on average,
the single fibre of the SDSS legacy spectra at $0.05\lesssim z\lesssim 0.1$
(R$\leq$2.2\,kpc), and a much more extended
aperture, probing out to R$\leq$1.5$R_{\rm eff}$.  We find similar fractions
of AGN (6/8 - \textit{NGC1093, NGC2639, NGC2906, NGC2916, NGC6394 and 
NGC7466}) and SF 
(5/7 - \textit{NGC5947, NGC2253, NGC6004, NGC2916,
NGC0160}) galaxies to reside in the GV, 
within 2.2\,kpc aperture. However, we
find evidence of the role of AGN in quenching, as AGN galaxies in a
twin system have greater D$_n$(4000) than their SF counterpart $90\%$
of the time, regardless of aperture size (Fig.~\ref{fig:Bimod_Min}).

The sample is divided into a set of groups depending on their location
in this diagram (i.e. whether they live on the RS, GV, or BC) finding
diverse sets of grouping.  We base this classification on spectra 
in the most central (1.5,\arcsec)
aperture as AGN activity is thought to impact more significantly the
immediate vicinity.  We grouped twins 1, 3 and 4 into group G1 as they
all have their AGN in the RS, while the SF counterpart resides in the
BC. Similarly, group G2 (twin 2 and 5) have their AGN in the RS but
the SF system also resides in the RS. Group G3 (twin 7 and 8) have
both AGN and SF in the GV. Finally, G4 (twin 5) has the AGN in the GV,
while the SF galaxies are in the BC.

The twin sample is compared with a general distribution of galaxies
from SDSS, with similar stellar mass.  All twin AGN galaxies were
selected in the CALIFA sample as type~2 Seyfert AGN, therefore, we
need to select the same type in the SDSS sample.  We 
separate type~1 and 2 AGN in
the Seyfert sample from SDSS.  This novel method makes use of the
equivalent width of the H$\alpha$ line in SF galaxies, to calibrate
the peak of H$\alpha$ emission in lines without a broad component.

Furthermore,  statistical tests show all twin galaxies have a 
relative difference within 3$\sigma$ and all
but one twin had difference within 1$\sigma$ for randomly selected
galaxies of only SF type or SF and AGN type (Fig.~\ref{fig:RD_D4k}).
Note, due to the small sample size, we do not find any strong 
conclusive results. However, we always find the relative variation 
to be positive, $\gtrsim 0$,
indicating, to a minor degree, a consistent physical difference 
caused by the AGN activity.
Therefore, this study should be treated more as a \textit{pilot study},
which motivates the extension of this methodology to a larger sample
to confirm, and draw a strong conclusion on whether such positive difference 
between ``twin'' samples is
due to the effect of the AGN on the evolution of the star formation
history of galaxies.

The analysis of a targeted set of line strengths showed a mixture of behaviour
regarding the stellar populations. These results were discussed in the
framework of two alternative hypothesis, one invoking the ``on-off'' switching
of the AGN, and the other assuming that while in some galaxies the 
AGN exerts the dominant form of quenching, in others quenching is not related to 
AGN activity - named ``no-AGN''. 
AGN galaxies generally have stronger 4000\AA\ break 
and [MgFe]$^\prime$ and weaker H$\delta_A$ within 1.5\arcsec and 2.2\,kpc
aperture -- indicating 
an older and more metal rich population. However, converting
these line strengths to SSP equivalent ages and metallicities
shows there is no clear trend in stellar age between AGN
and SF galaxies (Fig.~\ref{fig:Fid_SSP_Age}). 
In contrast to age, 70$\%$, $80\%$ and 70$\%$ of AGN
galaxies appear more metal rich within 1.5\arcsec, 2.2\,kpc and 
1.5\,R$_{\rm eff}$ (Fig.~\ref{fig:Fid_SSP_Met}),
respectively indicating even amongst ``twin'' sample AGN galaxies  
have a different formation time or star formation and chemical enrichment histories.
The assumption of different formation time would fit 
the ``on-off'' AGN hypothesis, while the different star formation and
 chemical evolution histories would support the ``no-AGN'' scenario. 

Finally, the presence of a bar may play an important role,
as we can rearrange the grouped twins into sets 
where groups 1 and 4 have a bar and groups 2 and 3 
are galaxies without bar. Groups 1 and 4 
feature consistent differences between 
AGN and SF galaxies, mostly supporting the no-AGN theory or 
suggesting a different formation time,
whereas groups 2 and 3, show more similarities
than differences between AGN and SF galaxies, supporting 
the ``on-off'' AGN switching mode hypothesis. The diversity of
the results in this sample shows the complex behaviour, thus no
definitive conclusion can be drawn. However, this work
provides a strong justification for a larger study that adopts the 
methodology implemented in this paper - with upcoming surveys
owing to the difficulty of finding ``twin'' galaxies - 
to study the effect of AGN on the evolution of star formation in galaxies.

\section{Acknowledgements}
The referee, Yago Ascasibar, is thanked for his constructive criticism.
JA is supported by the UK Science and Technology Facilities Council
(STFC).  IMC acknowledges the support of the 
Instituto de Astrof\'isica de Canarias via an Astrophysicist Resident fellowship.
We acknowledge support from the Spanish Ministry of Science,
Innovation and Universities (MCIU), through grant PID2019-104788GB-I00
(IF).  BG-L acknowledges support from the Spanish 
Ministry of Science, Innovation and Universities (MCIU), 
Agencia Estatal de Investigaci\'on (AEI), and the 
Fondo Europeo de Desarrollo Regional (EU-FEDER) under  projects  with  
references  AYA20155-68217-P  and PID2019-107010GB-100. 
CRA acknowledges financial support from the Spanish Ministry of 
Science, Innovation
and Universities (MCIU) under grant with reference RYC-2014-15779, 
from the European Union's Horizon
2020 research and innovation programme under Marie Sk\l odowska-Curie
grant agreement No 860744 (BiD4BESt), from the State
Research Agency (AEI-MCINN) of the Spanish MCIU under grants
"Feeding and feedback in active galaxies" with reference
PID2019-106027GB-C42, "Feeding, feedback and obscuration in active 
galaxies"
with reference AYA2016-76682-C3-2-P, and "Quantifying the impact of 
quasar feedback on galaxy evolution (QSOFEED)" with reference 
EUR2020-112266. CRA also acknowledges support from the 
Consejer\'ia de Econom\'ia, Conocimiento y Empleo del Gobierno de 
Canarias and the European Regional Development Fund (ERDF) under grant
with reference ProID2020010105 and from IAC project
P/301404, financed by the Ministry of Science and Innovation, through
the State Budget and by the Canary Islands Department of Economy,
Knowledge and Employment, through the Regional Budget of the 
Autonomous Community. This study uses data provided by the Calar Alto Legacy Integral 
Field Area (CALIFA) survey (http://califa.caha.es/).
Based on observations collected at the Centro Astron\'omico Hispano Alem\'an (CAHA) at Calar 
Alto, operated jointly by the Max-Planck-Institut f\"ur Astronomie and the Instituto de 
Astrof\'isica de Andaluc\'ia (CSIC).
Funding for SDSS-III has been provided by the Alfred P. Sloan
Foundation, the Participating Institutions, the National Science
Foundation, and the U.S. Department of Energy Office of Science. The
SDSS-III web site is http://www.sdss3.org/.

\section{Data Availability}
The data regarding SDSS (http://skyserver.sdss.org/dr16) and
CALIFA (https://califa.caha.es) data, used in this paper 
is publicly available at the respective websites. The data used for this 
project is available upon reasonable request.

\bibliographystyle{mnras}
\bibliography{TwinGV_Rev}

\appendix
\section{Type 1 and 2 AGN and SF galaxy spectra}
\label{sec:spec_sel}
\begin{figure}
    \centering
    \includegraphics[width=\linewidth]{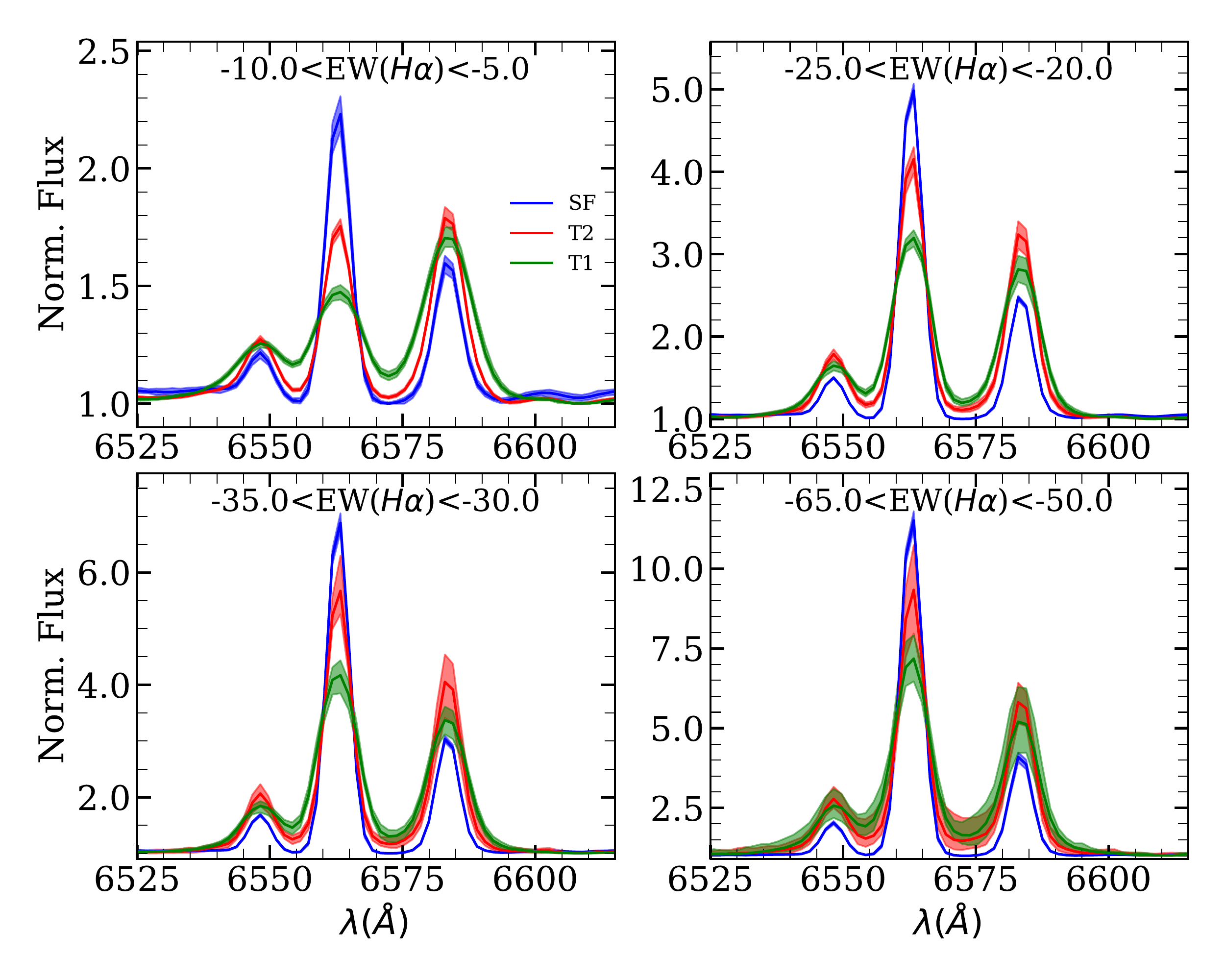}
    \caption{Stacked spectra of galaxies in the H$\alpha$+[N{\sc II}] region,
    classified as either star-forming (blue, SF), type~1 (green, T1) and type~2 (red, T2)
    AGN. The stacks corresponding to AGN (SF) galaxies 
    consist of 30\% (5\%) of the total galaxies in the EW(H$\alpha$) bins.}
    \label{fig:Gal_Spectra}
\end{figure}

We test here the validity of the approach followed in Sec.~\ref{sec:T2_AGN_Selection}
to further classify SDSS spectra flagged as Seyfert AGN (BPT flag 4 in
the {\sc galSpecExtra} catalogue) into type~1 and type~2. The criterion is
based on a comparison between line amplitude and equivalent width
(Fig.~\ref{fig:T2_sel}), so that the AGN sample that closely follows the
trend of star-forming galaxies (BPT flag 1) is supposed to be type~2
AGN, i.e. lacking a broad component.  Fig.~\ref{fig:Gal_Spectra} shows
the stacked spectra of SF, type~1 and 2 AGN for different bins of
EW(H$\alpha$) in blue, green and red, respectively (following the same colour scheme as Fig.~\ref{fig:T2_sel}). For each bin, we
make 50 stacks, consisting of 30$\%$ of type~1 and 2 AGN galaxies and
5\% of SF galaxies. We then plot the mean spectrum, where the errorbar
(filled colour) indicates the minimum and maximum flux in each stack
at a given wavelength. The spectra are normalised in the displayed
spectral window, dividing it by the minimum flux in each individual
stack.  The figure illustrates the expected trend, where galaxies
classified as type~1 have a clear signature of a broad component. At
the lowest EW(H$\alpha$), where we have the highest number of
galaxies, we find the cleanest separation, where the green line
features a wider H$\alpha$ line compared to both red and blue
lines. This is also evident at higher EW(H$\alpha$), but there are
fewer galaxies with a broad component (type~1) in this case.

\section{Age-Metallicity Degeneracy}
\label{sec:AB_Deg}

\begin{figure}
    \centering
    \includegraphics[width=\linewidth]{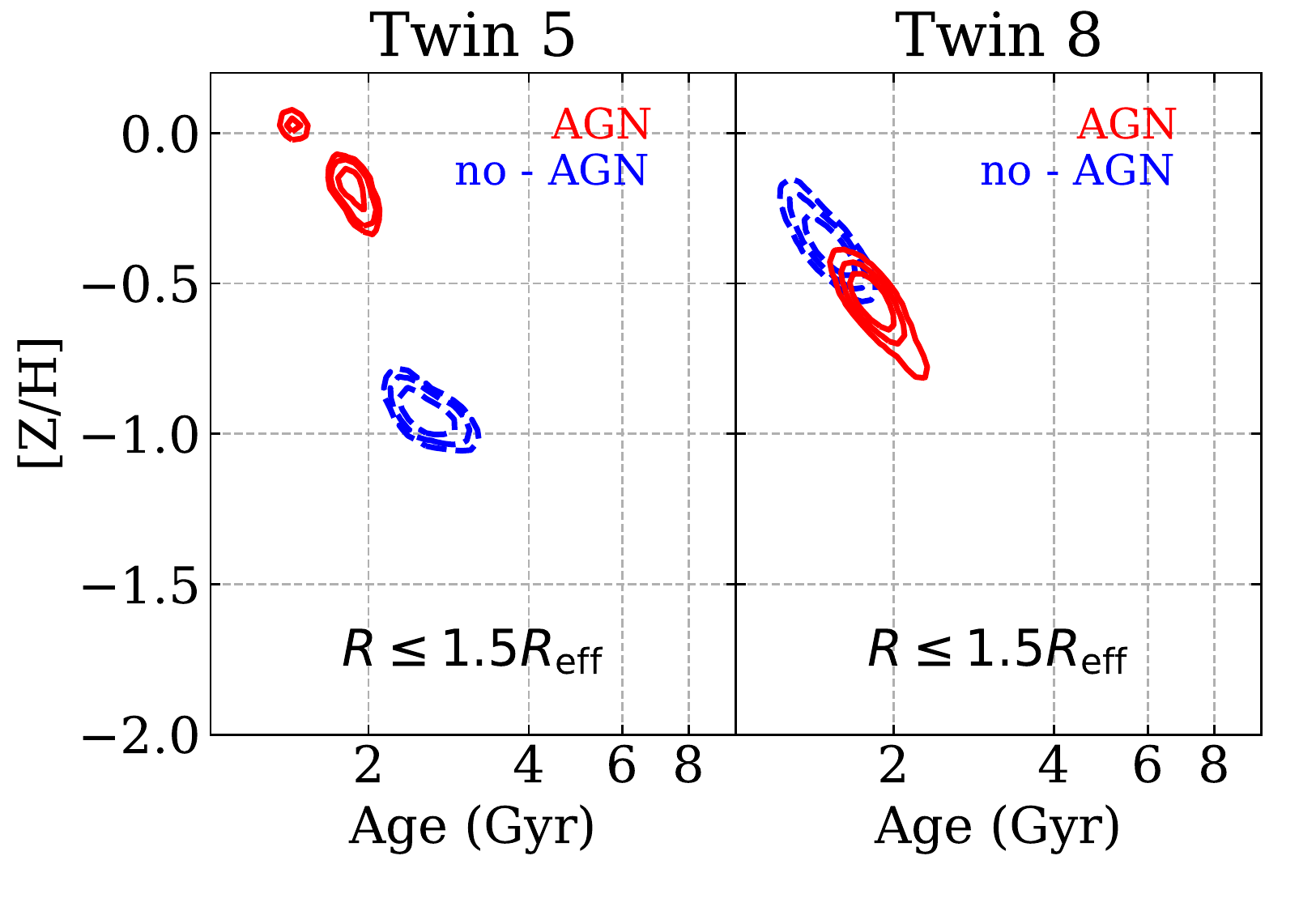}
    \caption{Confidence levels of the SSP-equivalent age and metallicity corresponding to
    galaxies in twin 8, using the largest (R$\leq$1.5\,R$_{\rm eff}$) aperture. The contours
    represent, from the inside-out the 1, 2 and 3 $\sigma$ levels.
    Solid red and dashed blue contours show the results for the AGN and 
    SF galaxy, respectively. }
    \label{fig:Con_FGal}
\end{figure}

The line strengths are all prone to the age-metallicity 
degeneracy \citep{Worthey:94}, therefore affecting the estimation
of our SSP equivalent ages and metallicities. This degeneracy
implies that the effect of an old stellar population can be
mimicked by a higher metallicity, producing very similar
colours or even line strengths \citep{ameg}. To break such 
degeneracy, we make use of a battery of line strengths
with different sensitivity to age and chemical composition.
Our set of indices comprises: D$_n$(4000),
H$\delta_A$, H$\gamma_A$, Mgb, Fe5270, Fe5335 and [MgFe]$^\prime$.
The spectra have a high S/N, which helps to break such degeneracy. 
Fig.~\ref{fig:Con_FGal} shows a bivariate plot with the confidence
levels for the SSP-equivalent age and metallicity  of twin 5 and 8.
The contours are shown -- from the inside out -- at the 1, 2, and 3\,$\sigma$
level. The solid red (dashed blue) contours correspond to the 
twin with (without) an AGN. 
A large overlap between these contours  would mean a substantial 
age-metallicity
degeneracy, so that the best-fit SSP parameters would be affected 
by it. We find no overlap between AGN and SF contours for 
twin 5, indicating that similar trends to twin 5 in 
R$\leq 1.5\,$R${\rm eff}$ should be independent of the 
age-metallicity degeneracy. Twin 8 shows some level of overlap 
between the contours, indicating there might be 
some level of degeneracy, however within 1$\sigma$
we find there to be no overlap in confidence levels, therefore 
indicating that the difference in age/metallicity is robust.
Furthermore, our $\chi^2$ analysis consists of artificially boosting
the uncertainty by adding, in quadrature,  5$\%$ of the index value due
to our models not being accurate enough to explain the observations.
Therefore, our confidence level should be tighter, thus breaking
the age-metallicity degeneracy, as we are unlikely to have large 
overlaps on the contours.


\section{Choice of apertures}
\label{Sec:aper_sec}
\begin{figure}
    \centering
    \includegraphics[width=0.9\linewidth]{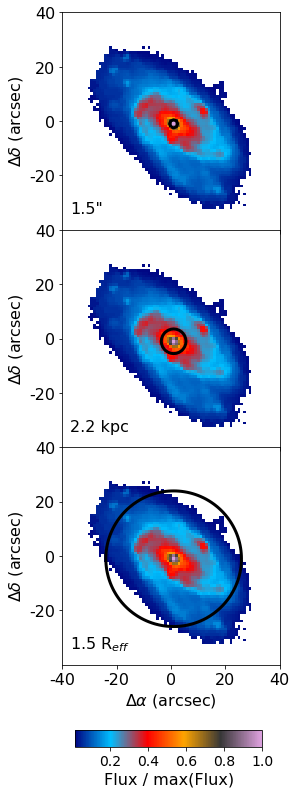}
    \caption{ Stellar flux map, UGC00005, of a data cube from the CALIFA survey. 
    \textit{Top, middle} and \textit{bottom:} panels show in black the most central
    radial aperture, R$\leq$1.5\arcsec, radial aperture to match the SDSS survey, 
    R$\leq2.2$\,kpc, 
    and maximum radial aperture, R$\leq 1.5$R$_{\rm eff}$, respectively.}
    \label{fig:apertures}
\end{figure}
In order to assess the effect of AGN activity on its host galaxy, we study 
the stellar population properties of galaxies at various apertures. 
Fig.~\ref{fig:apertures} shows the 3 apertures we have chosen to analyse the 
stellar populations, quoted in terms of radius R$\leq 1.5$\arcsec,
R$\leq 2.2\,$kpc and R$\leq 1.5$R$_{\rm eff}$. The spectra from individual 
spaxels, with S/N$\geq 3$, are summed up within each aperture,
leading to high signal to noise ratio, S/N$\geq 100$. Additionally, the spectra are brought to rest-frame
using the velocity maps computed in \cite{Ig:2020}.

\section{Galaxy SDSS images}
\label{Sec:gal_img}
\begin{figure}
\centering
\includegraphics[width=\linewidth, height=130mm]{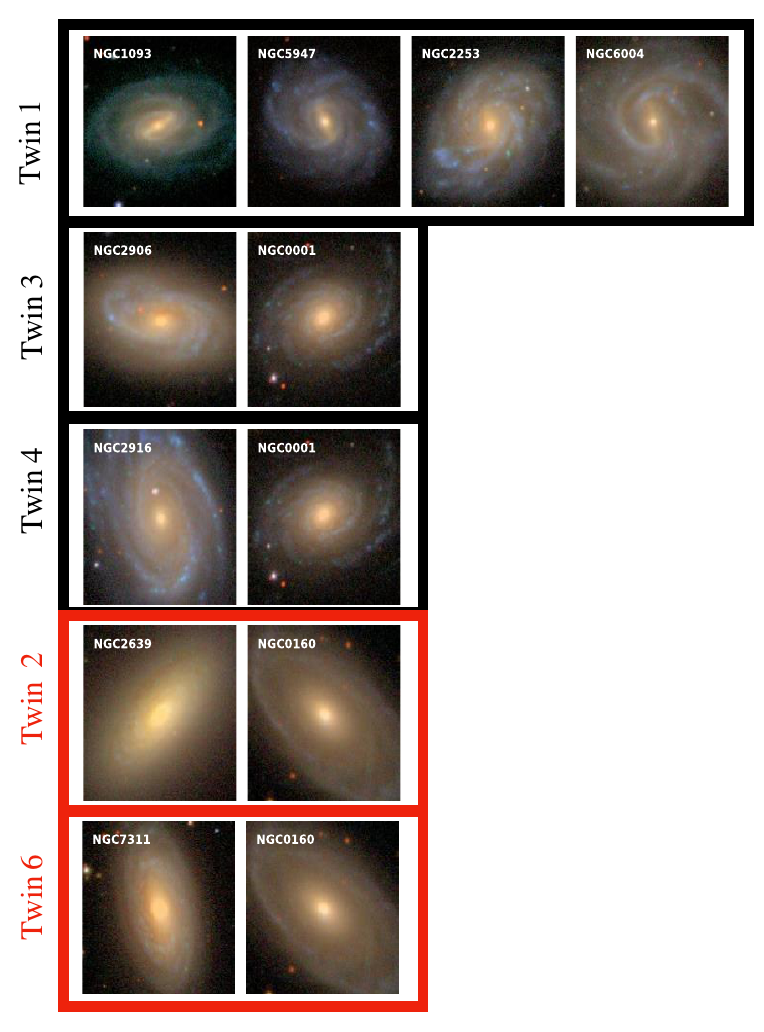}

\caption[]{Colour-composite SDSS images of the active galaxies (left column) and their 
corresponding non-active twin(s). Each image has a field of view of 
90\arcsec$\times$90\arcsec. North is up and east to the left. The 
frame colour indicates the group that each of the twin pair belongs to. 
Here, twins with black and red
frames indicate G1 and G2 galaxies, respectively.}
\label{fig:galaxy_images_1}
\end{figure}

Here we display stamp-like images, obtained from the CALIFA collaboration
using SDSS images,
of the various twins (Fig.~\ref{fig:galaxy_images_1}
and Fig.~\ref{fig:galaxy_images_2}). Each row 
corresponds to a twin pairing, where the left-most galaxy is an AGN, while 
the rest are SF. For the different groups identified 
in this paper, we have framed the respective twins accordingly. 
G1, G2, G3 and G4 twins are framed in black, red, blue and green, 
respectively.

\begin{figure}
\begin{center} 

\includegraphics[width=\linewidth, height=90mm]{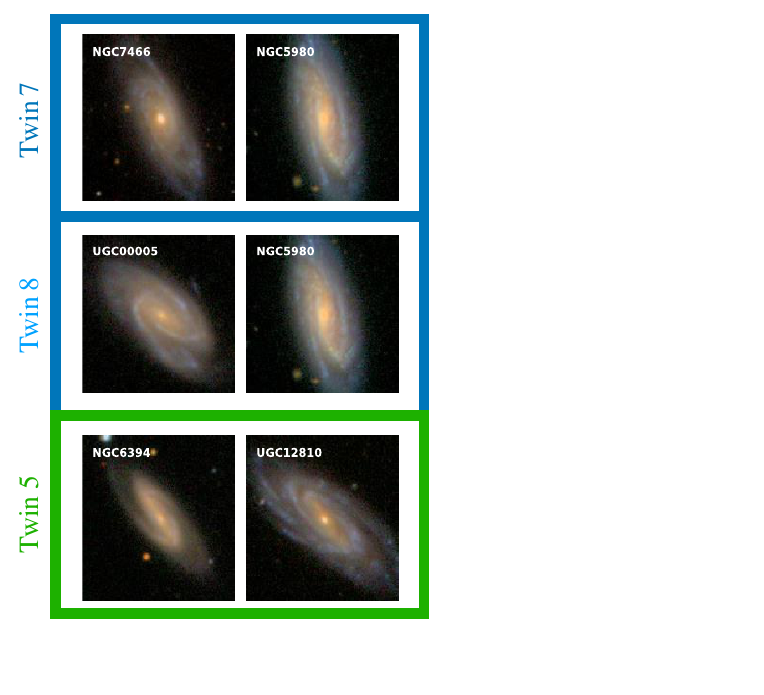}
\caption[]{Continuation of Fig. \ref{fig:galaxy_images_1}.The blue and 
green frames indicate groups G3 and G4, respectively.}

\label{fig:galaxy_images_2}
 \end{center}
\end{figure}
\label{lastpage}

\end{document}